\documentclass[twocolumn, pra, aps, longbibliography, superscriptaddress]{revtex4-2}
\usepackage{amssymb, amsmath, amsfonts}

\usepackage[]{inputenc}
\usepackage{color}
\usepackage{amstext}
\usepackage{enumitem}
\usepackage{graphicx, bm, palatino}
\usepackage[colorlinks=true, linkcolor=blue, urlcolor=blue, citecolor=blue, pdfusetitle]{hyperref}
\usepackage{cleveref}
\usepackage{microtype}

\usepackage[sc]{mathpazo}

\usepackage{cleveref}
\usepackage[dvipsnames]{xcolor}
\usepackage[caption=false]{subfig}

\usepackage{times}
\usepackage{bbm}

\DeclareMathOperator{\tr}{tr}


\begin{document}
\title{Speeding up thermalization and quantum state preparation through engineered quantum collisions}

\author{Sofia Sgroi}
\affiliation{Universit\`a degli Studi di Palermo, Dipartimento di Fisica e Chimica - Emilio Segr\`e, via Archirafi 36, I-90123 Palermo, Italy}

\author{Salvatore Lorenzo}
\affiliation{Universit\`a degli Studi di Palermo, Dipartimento di Fisica e Chimica - Emilio Segr\`e, via Archirafi 36, I-90123 Palermo, Italy}

\author{Luca Innocenti}
\affiliation{Universit\`a degli Studi di Palermo, Dipartimento di Fisica e Chimica - Emilio Segr\`e, via Archirafi 36, I-90123 Palermo, Italy}

\author{Paolo A. Erdman}
\affiliation{Universit\`a degli Studi di Palermo, Dipartimento di Fisica e Chimica - Emilio Segr\`e, via Archirafi 36, I-90123 Palermo, Italy}

\author{G. Massimo Palma}
\affiliation{Universit\`a degli Studi di Palermo, Dipartimento di Fisica e Chimica - Emilio Segr\`e, via Archirafi 36, I-90123 Palermo, Italy}

\author{Mauro Paternostro}
\affiliation{Universit\`a degli Studi di Palermo, Dipartimento di Fisica e Chimica - Emilio Segr\`e, via Archirafi 36, I-90123 Palermo, Italy}
\affiliation{Centre for Quantum Materials and Technologies, School of Mathematics and Physics, Queen's University Belfast, BT7 1NN, United Kingdom}

\begin{abstract}
We realize fast thermalization and state preparation of a single mode cavity field using a collision model-like approach, where a sequence of qubits or three level system ancillae, sequentially interacting with the field, is engineered with a genetic algorithm approach. In contrast to optimal control techniques, there is no time-dependent system Hamiltonian control deployed and, in contrast to reservoir engineering, the engineered full system-environment dynamics is optimized, and the target is not a steady state. We prove a significant speed up in thermalization --- for which we show that diagonal qubit ancilla states are sufficient --- and in the preparation of coherent states. We demonstrate the preparation of squeezed states and of highly non-Gaussian states. Our work offers a new alternative for fast preparation of cavity states that lays in between optimal Hamiltonian control and reservoir engineering, and gives further insights on the resources needed to realize or speed up the preparation of such states. 
\end{abstract}

\maketitle

\section{Introduction}\label{section:introduction}
The ability to prepare a quantum system in a desired state is a prerequisite for the realization of countless physical experiments and applications in quantum technologies. 
Moreover, even precise state preparation is of limited use if it requires very large preparation times; speeding up the process of state preparation is pivotal to design practical quantum technologies. To complicate the matter further, 
in many settings, one must not only reach the desired state, but also \textit{stabilize it} afterwards.

A possible approach to fast quantum state preparation is to initialize the system in a state that we can easily prepare and then perform a sequence of engineered gates~\cite{PhysRevA.83.032302}, or drive the system Hamiltonian suitably through techniques such as shortcut-to-adiabaticity~\cite{Berry_2009, RevModPhys.91.045001}, optimal control~\cite{Glaser2015, PRXQuantum.2.030203, wilhelm2020introductionoptimalcontrolquantum} or reinforcement learning~\cite{PhysRevX.8.031086, Zhang2019,Brown2021}. On the other end, stabilization to a target state can be achieved through reservoir engineering: the system is coupled to an environment that leads the former  to a target stationary state~\cite{PhysRevA.78.042307, PhysRevLett.86.4988, PhysRevLett.107.010402, PhysRevA.86.012114}. Such approach, however, guarantees convergence only asymptotically, i.e. for an infinitely long dynamics. It is thus meaningful to ask whether modulating the coupled system-environment dynamics can accelerate the process of state preparation. This will come at the cost of assuming a known initial system state but, once a desired proximity to the target is reached, we can stop modulating the dynamics, switching back to standard reservoir engineering and hence stabilizing the stationary state. A similar problem has already been explored in the context of thermalization and with a control limited to the system Hamiltonian. The ability to speed up thermalization of quantum systems is extremely useful in quantum thermodynamics, for example to increase the power output of quantum thermal engines. Under the appropriate conditions, it has been shown that it is possible to accelerate thermalization by coupling the system to a reservoir at the desired temperature and driving the system Hamiltonian \cite{PhysRevA.88.062326, PhysRevLett.122.250402, PhysRevA.101.052102, Kallush_2022, Pedram_2023}. Here we consider the opposite scenario, closer to reservoir engineering: we fix the Hamiltonian of the system and we focus on designing the environment.

Controlling the dynamics of the reservoir in order to prepare a smaller system in the desired state might seem like a daunting task and a complication from the original problem. While, in general, that is the case, we can suitably choose the environment so that such control might be achievable. In particular, we borrow from collisional models (CM) ~\cite{CICCARELLO20221, ciccarello2017collision, alicki2007quantum, PhysRevLett.88.097905} and we consider the case where the environment can be thought of as constituted from smaller subsystems, or ancillae, that sequentially interact with the system individually and each of which is discarded before the next interaction. Reservoir engineering can, in principle, be achieved through the tool provided by CMs via homogenization \cite{ziman2001quantumhomogenization}: by preparing a suitable number of ancillae -- all isodimensional to the system -- in the very same target state to reach and engineering a suitable system-ancilla interaction, one would be able to lead the system to such  target state. This is, however, not particularly useful for state preparation, as it requires us to already be able to prepare the ancillae in the desired state. Instead, we consider the case where the dimension of each ancilla is smaller than that of the system. In this case, we require the ability to prepare only smaller systems in a suitable state, making the approach meaningful for the problem of quantum state preparation.

To this end, we consider the case of a harmonic oscillator interacting sequentially with two- or three-level ancillae. This scenario is similar to that of a ``micromaser"~\cite{PhysRevLett.64.2783}, where atoms are sequentially injected into an optical cavity and interact with its field. Such an analogy not only serves an intuitive illustration of our methodology but also provides a concrete example of a physical platform for implementation. 
The problem of stabilization in a similar scenario has already been addressed for some cavity states~\cite{PhysRevLett.98.240401, PhysRevLett.107.010402, PhysRevA.86.012114, Miao_2017}. Here, we study the problem of fast state preparation: 
we initialize the state of each ancilla so that the sequential interactions lead the oscillator as close as possible to the target state, in the desired timeframe.
The optimal sequence of ancilla states is found using either a simple genetic algorithm~\cite{Goldberg1988, holland1975adaptation} --- fixing the number of collisions and all the other parameters of the dynamics --- or a specifically designed genetic algorithm that allows us to optimize the number of collisions within a given time together with the state of each ancilla. 
Genetic and evolutionary algorithms~\cite{eiben2005evolutionarycomputing}, have been previously applied in the context of quantum control to quantum circuit design~\cite{lukac2002evolving, rubinstein2001evolving, williams1998automated, sunkel2023ga4qcogeneticalgorithmquantum} and optimal control via Hamiltonian driving~\cite{Brown_2023, PALITTAPONGARNPIM2017116, PhysRevA.90.032310}.
Genetic algorithms were also applied in the context of CMs~\cite{Chisholm_2021}, but in a completely different context, and involving only the optimization over a few parameters, rather than to the entire ancilla sequence.

Using our approach, we are able to speed up thermalization and preparation of coherent states using a linear system-ancilla coupling. Fast preparation of squeezed states is achieved using a non-linear coupling. Preparation of non-Gaussian states is also demonstrated in both scenarios. Moreover, our analysis provides us with interesting insights on the CM considered and on the resources required to speed up thermalization. In particular, we show that the trace distance between the final state and the target depends non-monotonically on the collision time and that the optimal collision time depends on the temperature of the target state. Even more interestingly, while necessary for preparation of other states, coherences on the ancillae are not required to speed up the thermalization process, nor do they seem to help accelerating such process at all.

The paper is organized as follows. In Sec.~\ref{section:physicalmodel} we introduce the system and the process of interest in details. In Section~\ref{section:geneticalgorithms} we describe the genetic algorithms used for the optimization. Results for different target states are presented in Sec.~\ref{section:results}. In particular, we demonstrate the preparation of thermal, coherent and squeezed states  (Secs.~\ref{section:results:thermalstates}, \ref{section:results:coherentstates}, and \ref{section:results:squeezedstates}, respectively), and we find the maximally non-Gaussian state achievable with our protocol in Sec.~\ref{section:results:nongaussianstates}. Finally, our conclusions are discussed in Section~\ref{section:conclusions}.

\section{Physical Model}\label{section:physicalmodel}

\begin{figure}
\centering\includegraphics[width=\columnwidth]{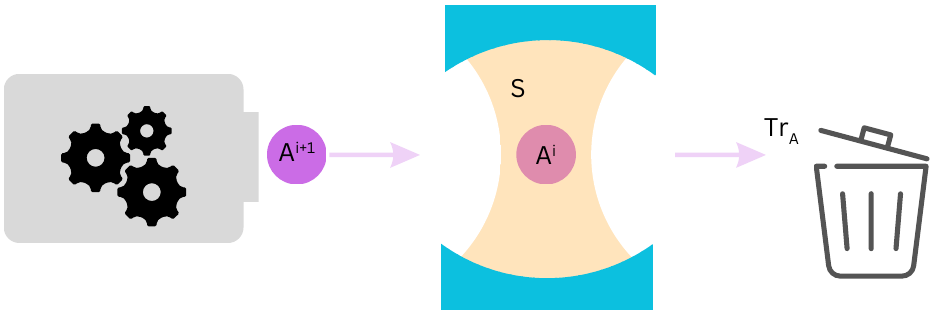}
\caption{Sketch of the maser-like physical process described in Sec.~\ref{section:physicalmodel}. A sequential stream of ancillae individually interact with a cavity before being discarded. In CMs, the ancillae are all prepared in a specific state depending on the environment under consideration. Here, the ancillary states are thoroughly engineered to lead the cavity state toward a desired target.}\label{Figure:maser}
\end{figure}

Let us consider a single-mode cavity field at frequency $\omega_C$. Its Hamiltonian can be written as
\begin{equation}
    H_S = \hbar \omega_C \bigg(a^{\dagger}a + \frac{1}{2}\bigg),
\end{equation}
where $a$ is the annihilation operator. The dynamics under consideration consists of a sequence of system-ancilla interaction steps, or "collisions" (a sketch of the process is illustrated in Fig.~\ref{Figure:maser}). At each step, the system and the ancilla, initially in a product state, first evolve unitarily for a collision time $t_c$, according to the Hamiltonian
\begin{equation}
    H = H_S + H_A + H_I,
\end{equation}
where $H_A$ is the ancilla Hamiltonian and $H_I$ the coupling Hamiltonian. All of these terms, wether optimized or not, will be kept constant during the evolution. Next, the ancilla state $\rho_A$ is discarded by tracing out the ancilla's degrees of freedom and the reduced system state $\rho_S$ is kept as the next initial system state. We consider no time delay in between collisions. Concisely, we can write the state of the system after $i$ collisions as
\begin{equation}\label{equation:collision}
    \rho_S^{i+1} = \tr_A\{U(\rho_S^{i}\otimes\rho_A^{i})U^{\dagger}\},
\end{equation}
where $U = \exp(- i H t_c /\hbar)$ is the propagator and each of the ancillary states $\rho_A^i$ is prepared independently. The total evolution time can be written as $T = n t_c$, where $n$ is the total number of collisions and $t_c$ is the interaction time for each collision. We will consider a coupling Hamiltonian of the form
\begin{equation}
    H_I = H_l + H_{nl},
\end{equation}
where $H_l$ and $H_{nl}$ are a linear --- single-photon --- and a nonlinear --- two-photons --- coupling, respectively.
The nonlinear coupling is introduced only as a last resource and a linear interaction will suffice for most of the cases considered throughout this work. However, we will show that, within our setting, preparation of squeezed state is reasonably achievable only by including $H_{nl}$. 

Unless otherwise specified, we will consider qubit ancillae, with Hamiltonian
\begin{equation}
    H_A = \frac{\hbar \omega_A}{2} \sigma_Z,
\end{equation}
where $\sigma_Z$ is the $Z$-Pauli matrix.
In these cases, the linear interaction term will be given by the Jaynes-Cummings Hamiltonian \cite{raimond2006exploring}
\begin{equation}\label{equation:linearIqubit}
    H_l = \hbar g_l (a\sigma_+ + a^{\dagger}\sigma_-)
\end{equation}
while the nonlinear interaction term will be of the form
\begin{equation}\label{equation:nonlinearIqubit}
    H_{nl} = \hbar g_{nl} (a^2\sigma_+ + a^{\dagger2}\sigma_-).
\end{equation}

We will also occasionally consider three-level system ancillae. In this case, write the ground state of the ancilla as $|0\rangle_A$ and set its energy to zero. The ancilla Hamiltonian is then written as
\begin{equation}
    H_A = \hbar \omega_{1}|1\rangle\langle 1|_A + \hbar\omega_{2}|2\rangle\langle 2|_A,
\end{equation}
where $|1\rangle_A$ and $|2\rangle_A$ are the first and the second excited states of the three-level system, respectively, and $\omega_{1}<\omega_{2}$. We consider
\begin{equation}
    H_l = \hbar g_{l1} a|0\rangle\langle 1|_A + \hbar g_{l2} a|1\rangle\langle 2|_A + h.c,
\end{equation}
and
\begin{equation}
    H_{nl} = \hbar g_{nl} a^2|0\rangle_A\langle 2|_A + h.c.,
\end{equation}
where $h.c.$ stands for Hermitian conjugate.

\section{Genetic Algorithms}\label{section:geneticalgorithms}
In this Section, we discuss the genetic algorithms (GA) used to optimize the ancilla state sequence and the other parameters. 

Let us start by discussing the generic structure of the genetic algorithm we consider~\cite{KONAK2006992, holland1975adaptation}. First, we define a \textit{Fitness} function that we aim to maximize. A set of $N$ candidate solutions is randomly generated, each encoded by a string of numbers, and their \textit{Fitness} evaluated. The following steps are then repeated for a given number of iterations, or until a desired convergence threshold:
\begin{itemize}
    \item[-]{\textbf{Selection}: among the population, a copy of the best $M$ solutions is passed unchanged to the next generation, i.e. for the next iteration of the algorithm.
    Then, $\frac{N-M}{2}$ couples of solutions are randomly selected from the current population with a likelihood dependent on their \textit{Fitness}.}
    \item[-]{\textbf{Crossover}: for each selected couple generate two offspring solutions to include in the population of the next generation. These are determined via some predefined crossover operation on the two parent solutions that combines some of the information from both.}
    \item[-]{\textbf{Mutation}: randomly mutate the offspring solutions with a certain probability.}
    \item[-]{\textbf{Fitness evaluation}: the \textit{Fitness} of the new solutions is evaluated.}
\end{itemize}

A simple flowchart of the algorithm described above can be found in Fig. \ref{Figure:flowchart}.

For this work, unless specified, we define the \textit{Fitness} function as the the opposite of the trace distance \cite{Nielsen_Chuang_2010} between the final state of the system $\rho^n_S$ and the desired target state $\rho^{target}$, i.e.
\begin{equation}\label{equation:trDfit}
    J (\rho^n_S, \rho^{target}) = -\frac{||\rho^n_S-\rho_{target}||}{2},
\end{equation}
where $||\rho|| = \tr(\sqrt{\rho^\dagger \rho})$. Here, $\rho^n_S$ is evaluated by iterating Eq.~\eqref{equation:collision} $n$ times. 

\begin{figure}
\centering\includegraphics[width=\columnwidth]{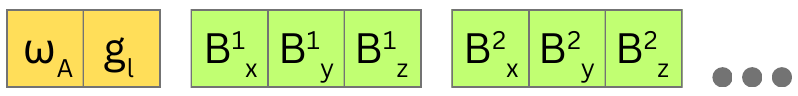}
\caption{Example of candidate solution as described in Section~\ref{section:geneticalgorithms}. In this example, we consider qubit ancillae interacting linearly with the cavity. The elements in the yellow cells are the parameters fixed during the dynamics (in this example, the ancilla frequency and the coupling constant), while the green cells contain a parametrization of the ancilla state at each collision (in this case, the components of the Boch vectors). Here $\lambda_0=2$ and $\lambda_d=3$.}\label{Figure:genexample}
\end{figure}

A candidate solution $x$ is encoded as a list of real numbers $x_i$ (an example is illustrated in Fig.~\ref{Figure:genexample}). The total length of the list is $L = \lambda_0 + n\lambda_d$, where the first $\lambda_0$ values represent the parameters to be optimized which are kept constant during the dynamics (e.g. the ancilla frequencies, the coupling constants) while $\lambda_d$ is the number of parameters required to encode the ancilla state before a collision. Each of these values is bounded to a specific range and rescaled so that $x_i\in[0,1]$. Throughout this work, we will consider three different scenarios: qubits in a diagonal state are defined by an inverse temperature $\beta^i_A$, qubits in a generic state are parametrized using the component of a Bloch vector \cite{Nielsen_Chuang_2010}, while thre-level system ancillae are specified using the SU(3) Euler angles representation \cite{ToddTilma_2002, ToddTilma_2002_2, byrd1997geometrysu3}. We refer the reader to Appendix~\ref{appendix1} for details.

Solutions are chosen for Crossover and Mutation using \textit{Tournament selection} i.e. randomly extracting $K$ solutions from the population and selecting the best among them. The specifics of the Crossover and Mutation operations depend on wether we fix $L$, i.e. the number of collisions, or not.

\subsection{Fixed length}\label{section:geneticalgorithms:fixedlength}
We apply this approach whenever we want to fix the collision time $t_c$ and the number of collisions $n$ a priori.

Given that we are dealing with continuous variable $x_i\in[0,1]$, Crossover is implemented by linear combinations \cite{haupt2004practical}
\begin{equation}\label{equation:crossover}
    x^{a, b} = \gamma x^{A,B} + (1-\gamma)x^{B,A},
\end{equation}
where $x^{A,B}$ are the parent solutions, $x^{a,b}$ are the offspring solutions and $\gamma\in[0,1]$ is a random variable extracted from a uniform distribution.

Mutation consists in scanning the solution and changing $x_i$ to a new random value in the same range, with a probability $p_\mu = \mu/L$, where the mutation factor $\mu$ is fixed as a hyperparameter for the optimization.

\subsection{Variable length}\label{section:geneticalgorithms:variablelength}
We apply this approach when we want to fix the total time $T$ of the evolution but optimize $t_c$ and $n$. In this case, $L$ is optimized by the algorithm and $n = \frac{L-\lambda_0}{\lambda_d}$ and $t_c = \frac{T}{n}$ are computed for each solution before evaluating the \textit{Fitness}.

\begin{figure*}[t]
\centering
 \centering\includegraphics[width=2\columnwidth]{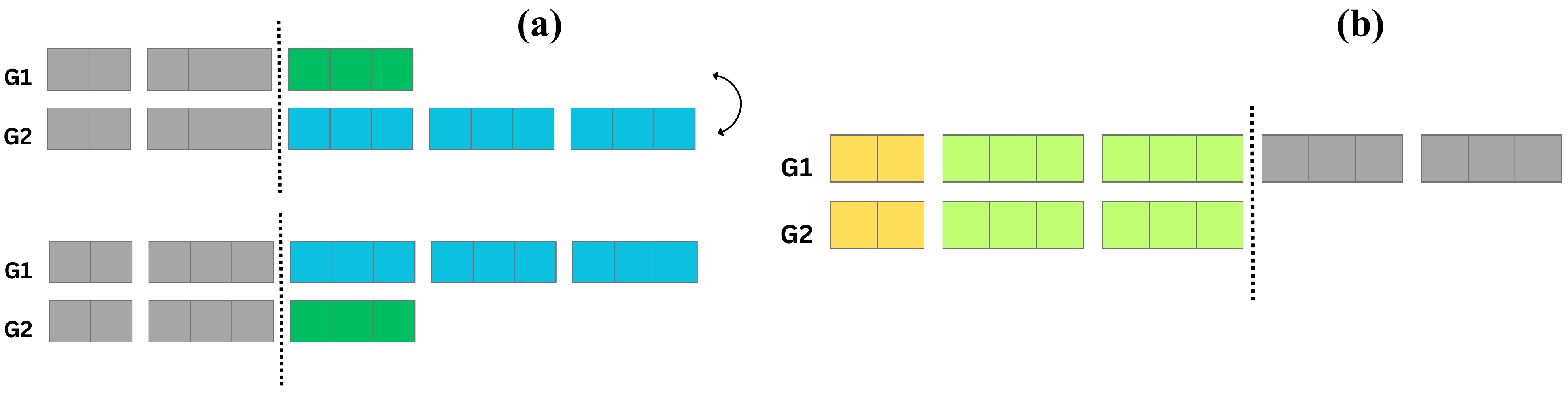} 
\caption{Sketch of the \textit{Crossover} operation described in Section~\ref{section:geneticalgorithms:variablelength}. \textbf{(a)} Each of the two parent solutions is split in two parts at a chosen, shared, collision. All the portion of the strings after such point (non-gray cells) are swapped between the two solutions. \textbf{(b)} Eq.\eqref{equation:crossover} is applied only to the shared components of the strings (non-gray cells).}\label{Figure:crossover}
\end{figure*}

Crossover consists in two steps. First, a random integer number $n_r\in[1,n_l]$ is extracted, where $n_l$ is the largest number of collisions among the two parent solutions. Each solution is the split at $\lambda_0 + \lambda_d\min(n, n_l)$ into two strings (one of the two string might be empty depending on $L, n_l$) which are are swapped between the two parent solutions. This gives us two temporary solutions. Then, we apply Eq.~\eqref{equation:crossover} to the shared components of the temporary solutions, to get the offspring (see Fig.~\ref{Figure:crossover} for a sketch of the process).

Mutation also consists in two steps: with a probability $\bar{p}$, a new collision, consisting of new random $\lambda_d$ elements, is added to the string. With the same probability, a collision is removed. Then, the solutions is scanned so that, with a probability $p_\mu = \mu/L$, each component $x_i$ might be changed to new random value in the same range.

The number of collisions is bounded: $n \in \{1, 2, ... , n_{max}-1, n_{max}\}$, where $n_{max}$ is fixed as a hyperparameter of the optimization.

\section{Results}\label{section:results}
In this Section, we present and discuss some numerical results following from the application of the approach proposed in Section~\ref{section:geneticalgorithms} to the model described in Section~\ref{section:physicalmodel} to prepare different cavity states. Throughout this Section, we fix $\omega_C=1$ and $\hbar=1$. Numerical and algorithm hyperparameter details can be found in Appendix~\ref{appendix2}.

\subsection{Thermal states}\label{section:results:thermalstates}
We start from the problem of thermalization. In this case, our goal is to prepare the cavity in a thermal state
\begin{equation}
    \rho^{target} = \frac{e^{-\beta H_S}}{\tr(e^{-\beta H_S})}.
\end{equation}
We consider resonant qubit ancillae and linear coupling, i.e. $\omega_A = \omega_C$ and $H_I$ is described by Eq.~\eqref{equation:linearIqubit} where, in particular, we fix $g=1$. We focus on the algorithm described in Section~\ref{section:geneticalgorithms:fixedlength} and we only optimize the ancilla state sequence for a given collision time and number of collisions. We start from generic qubit states.

\begin{figure}[b]
\centering\includegraphics[width=\columnwidth]{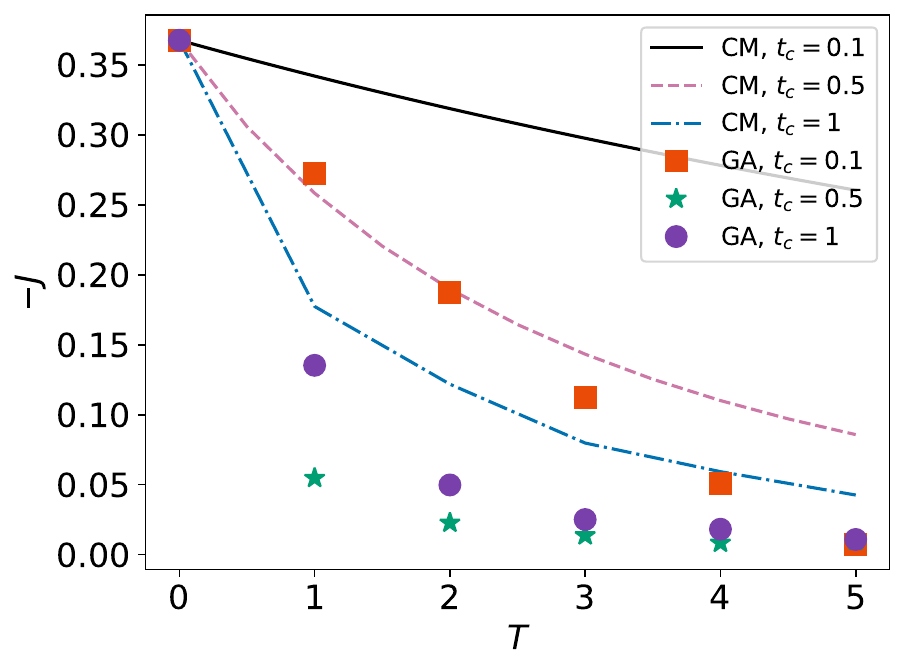}
\caption{Trace distance between the state of the system at a time $T$ and a thermal state at inverse temperature $\beta=1$ for collision model-like processes in which the ancillae are all prepared in a thermal state at inverse temperature $\beta_A=1$ (solid curves) and for the optimal solutions found with the genetic algorithm approach (dots), for different collision times. Each point is optimized at a different total evolution time $T$. Notice how the optimized solutions lead the system to the target much faster than the corresponding collision model-like thermalization process. Moreover, the speed-up is a non-monotonic function of the collision time.}\label{Figure:thermalization1}
\end{figure}

Let us first consider the problem of accelerating heating, which is known to be achievable with optimal control even in the adiabatic-master equation regime, in contrast to cooling \cite{PhysRevA.88.062326}. We initially prepared the cavity in the vacuum state and we tried to reach a thermal state with inverse temperature $\beta=1$. In principle, such state can be achieved by letting the system interact with ancillae all prepared in a thermal state at the desired inverse temperature $\beta$. Hence, we compared the performance of the optimized ancilla state sequence with this collision model-like dynamics for different collision times $t_c = 0.1, 0.5, 1$. Results are shown in Fig.~\ref{Figure:thermalization1}. The reason for such large collision times is both numerical and practical: decreasing the collision time would increase the number of necessary collisions i.e. the simulation time and the number of parameters to optimize while, at the same time, would require a hypothetical experimentalist to prepare a larger number of qubits in a specific state. However, notice that while this makes the system far from simulating a CM, and hence far from the corresponding master equation that would describe a more natural thermalization process, $J (\rho^i_S, \rho^{target})$ seems to be, at least in the regime considered, a monotonically decreasing function of time i.e. it would always outperform a master equation. Interestingly, however, for the optimized sequence this ceases to be the case: there seems to be an optimal non-obvious collision time, in this case $t_c=0.5$. The process with optimized ancillae always significantly outperforms the corresponding collision model-like process. For $t_c=0.5$, a small number of collisions appear to be sufficient to prepare the system in the target state fairly accurately: $J (\rho^{i=6}_S, \rho^{target})\approx -0.023$. At the same time ($T = 3$), a natural relaxation process would have only slightly moved the system from its initial state, in comparison. We can hence conclude that the desired thermalization process can be significantly accelerated in this scenario.

\begin{figure}[b]
\centering
\textbf{(a)}\\
 \centering\includegraphics[width=\columnwidth]{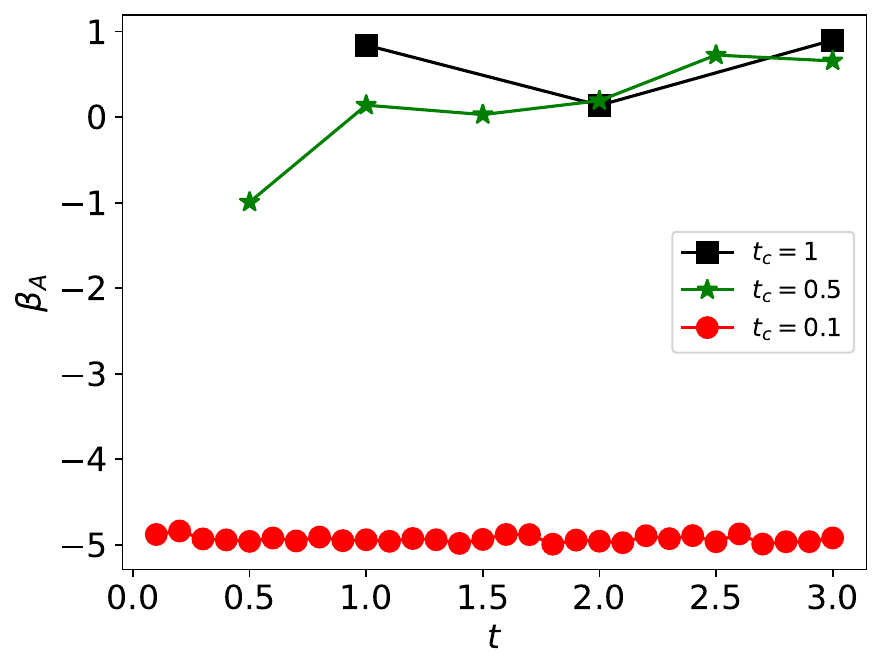}\\
 \textbf{(b)}\\
 \includegraphics[width=\columnwidth]{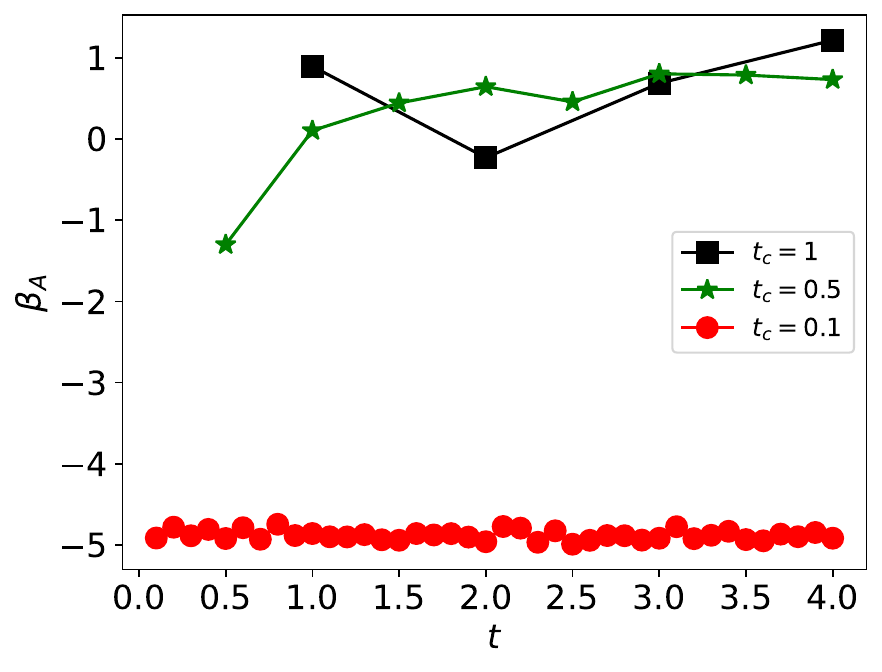}
\caption{Optimal ancilla inverse temperature sequence for different collision times for \textbf{(a)} $T=3$ and \textbf{(b)} $T=4$. The maximum ancialla generalized temperature achievable considered is for $\beta_A=-5$. Only for $t_c=0.1$, the best strategy is to heat the system as much as possible, i.e. send all the ancillae in the excited state.}\label{Figure:thermalization2}
\end{figure}

\begin{figure}
\centering\includegraphics[width=\columnwidth]{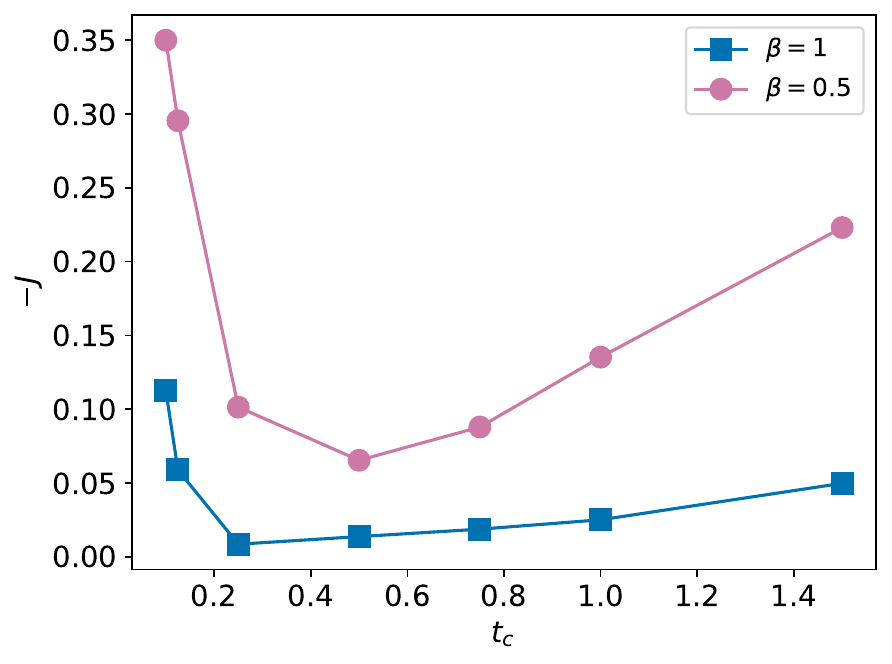}
\caption{Minimum trace distance between the final system state and the target for the optimal solution at different collision times and for two different target temperatures. The number of collision $n$ for each optimization is chosen so that the total time is $T=nt_c=3$. Notice that the optimal collision time depends on the target temperature. The trace distance for $\beta=0.5$ can be reduced by increasing the evolution time: for example, for $T=10$ and $t_c = 0.5$, $-J=0.036$.}\label{Figure:thermalization3}
\end{figure}

Given that we are considering the preparation of thermal state, it is interesting to ask whether the speed-up of the thermalization process at the desired temperature results from the right sequence of ancillae temperatures or the ancillae coherences contribute to such process. This question is non-trivial, as coherences in the ancillae are known to provide a system-environment energy exchange contribution akin to an effective work in the thermodynamics of collision models, at least for small coherences \cite{PhysRevLett.123.140601}. Hence, such coherences might, in principle, play a role in accelerating the system thermalization. This is, however, not the case here. Optimizing the ancilla state sequence by only considering diagonal ancillae does not, in fact, decrease the performance of the approach, as shown in Table~\ref{tab:bloch_vs_gibbs}. Examples of such optimized ancilla state sequence are shown in Fig.~\ref{Figure:thermalization2}, where each ancilla state is solely described by a generalized inverse temperature (we allow negative temperatures). The results both justify the use of our methodology and explain why  there exists an optimal collision time: as one can see from the plots, an optimal, non-trivial strategy exists and such strategy depends on both the collision time and the number of collisions. Only when the values of these two variables are not well suited to guarantee a reasonable proximity of the final state to the target thermal state, the best strategy becomes to "heat the system as much as possible". Moreover, the optimal collision time depends on the desired target state, as shown in Fig.~\ref{Figure:thermalization3}, which justifies the use of the algorithm described in Section~\ref{section:geneticalgorithms:variablelength} whenever we want to prepare the system in the desired target state within a given time $T$.

\begin{table}[b]
\begin{tabular}{  c |c  }
  $t_c$ & $-\Delta J$\\ [0.5ex]
 \hline
 \hline
 $1.0$ & $0.0038$ \\
 \hline
 $0.5$ & $0.0036$ \\
 \hline
 $0.1$ & $0.0106$ \\ [0.5ex]
\end{tabular}\caption{Maximum difference in the trace distance with the target of the optimal solutions found with generic and thermal ancillae for $T=1,2,3,4,5$. Thermal-ancillae solutions always slightly outperform the corresponding generic-ancillae solution, probably due to the fact that the optimization is significantly simpler (the optimization for thermal states involves only a third of the parameters for generic qubit states).}\label{tab:bloch_vs_gibbs}
\end{table}

\begin{figure}
\centering
\textbf{(a)}\\
 \centering\includegraphics[width=\columnwidth]{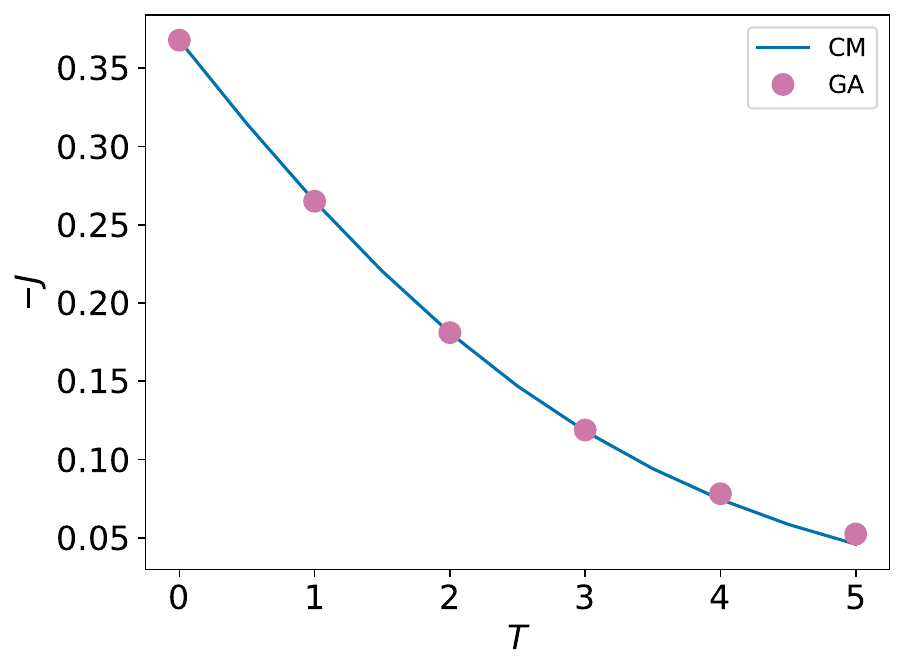}\\
 \textbf{(b)}\\
 \includegraphics[width=\columnwidth]{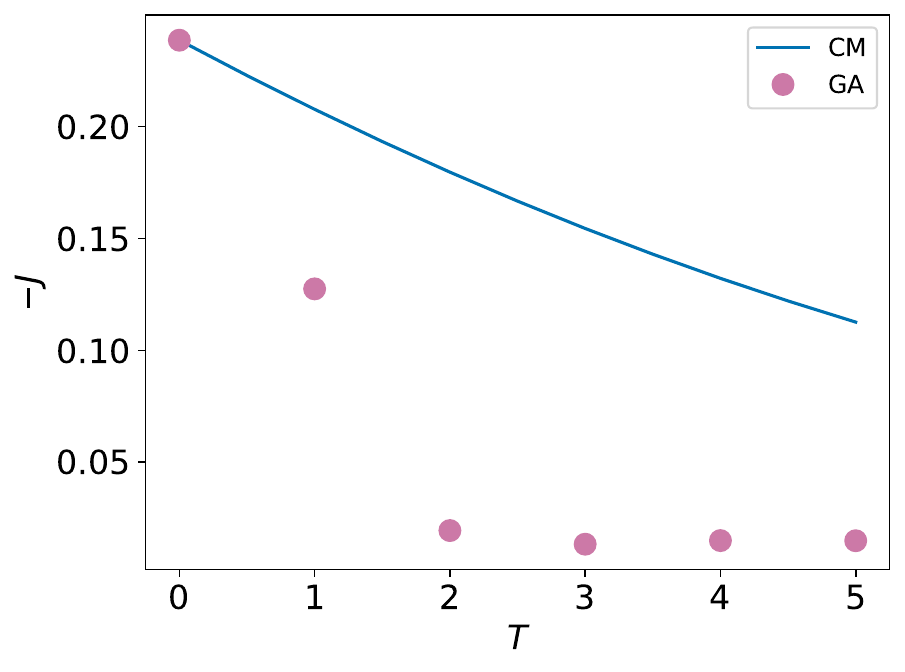}
\caption{Comparison between the trace distance between the final system state and the target for the collision model-like process (solid line) and the optimized ancilla sequence (dots) when \textbf{(a)} the the system is initialized at $\beta=1$ and the target is the void state and \textbf{(b)} the the system is initialized at $\beta=0.5$ and the target is a thermal state at $\beta=1$. In both cases, $t_c=0.5$.}\label{Figure:thermalization4}
\end{figure}

\begin{figure}
\centering\includegraphics[width=\columnwidth]{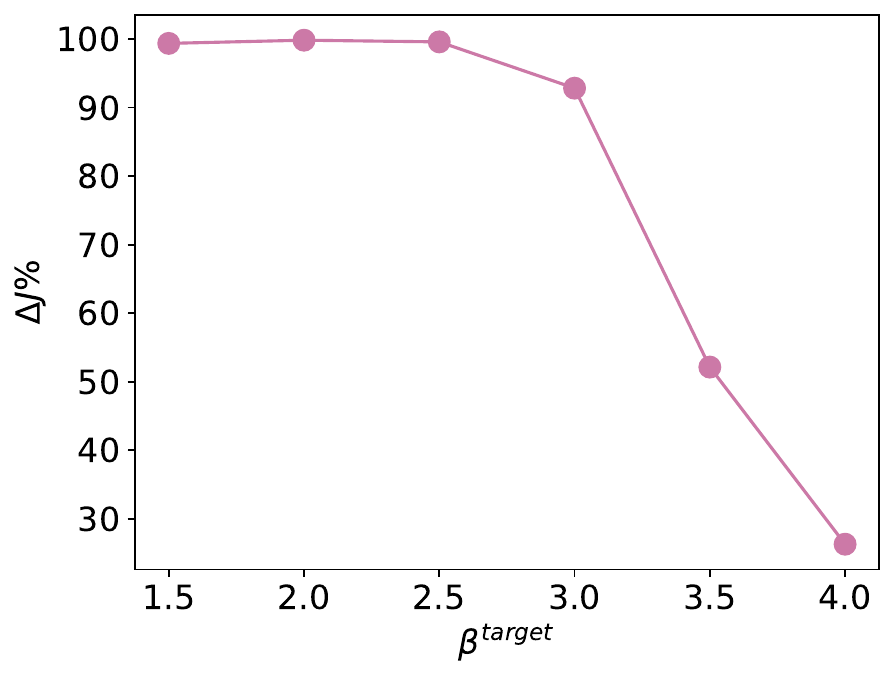}
\caption{Percentual final state-target trace distance reduction in using the optimized ancilla state sequence vs the collision model-like dynamics, as a function of the target inverse temperature. The state is initialized at $\beta=1$ and $\Delta J\%=-100\frac{J_{GA}-J_{CM}}{J_{CM}}$, where $J_{CM}$ and $J_{GA}$ are given by Eq.~\eqref{equation:trDfit} for $\rho_S^n$ obtained with the two different approaches.}\label{Figure:thermalization5}
\end{figure}

Finally, let us consider cooling. In contrast to heating, the ability to speed up the cooling down of the system with our approach seems to depend on the temperature of the target state. In particular, there is no advantage in optimizing the ancilla sequence for the reverse of the process described at the beginning of this subsection, i.e. when going from a thermal state with $\beta=1$ to the ground state. On the other end, speeding up the cooling process is possible if we consider non-zero temperatures. We report an example in Fig.~\ref{Figure:thermalization4}, where we initialized the system at $\beta=0.5$ and we moved towards $\beta=1$. Moreover, once again, coherences in the ancillae are not required, nor they seem to provide any advantage. As shown in Fig.~\ref{Figure:thermalization5}, the speed-up seems to be more pronounced the warmer is the target state. Nonetheless, the approach still provides some degree of acceleration of the thermalization process for low, even if not strictly zero, target state temperatures.

\subsection{Engineering coherent states}\label{section:results:coherentstates}
Let us now consider preparation of coherent states \cite{raimond2006exploring} of type
\begin{equation}\label{equation:coherentstates}
    \rho^{target} = D(\alpha)|0\rangle \langle 0| D (\alpha)^\dagger,
\end{equation}
where $D(\alpha)=e^{\alpha a^\dagger - \alpha^*a}$ is the displacement operator and $|0\rangle$ is the void state of the cavity. Once again, we consider linear system-ancilla interactions with $g=1$.

\begin{figure}
\centering\includegraphics[width=\columnwidth]{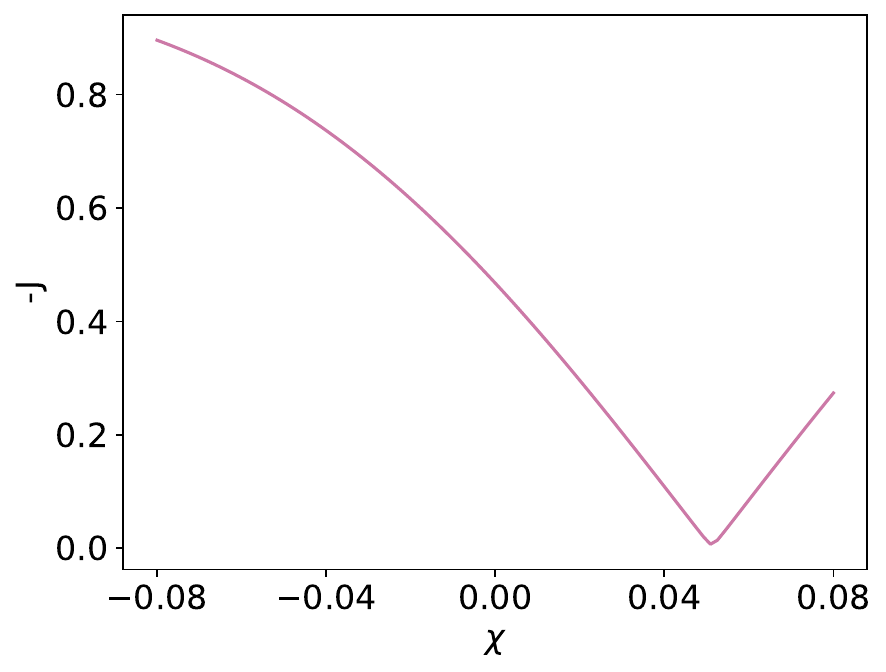}
\caption{Trace distance between a coherent state with $\alpha=0.5$ and the final state of the system, at a finite time $T=10$, obtained via a collision model-like process ($t_c=0.01$) with ancillae given by Eq.~\eqref{equation:ancillacoherences} with $\beta=5$, as a function of $\chi$.}\label{Figure:coherent1}
\end{figure}

\begin{figure}
\centering
\textbf{(a)}\\
 \centering\includegraphics[width=\columnwidth]{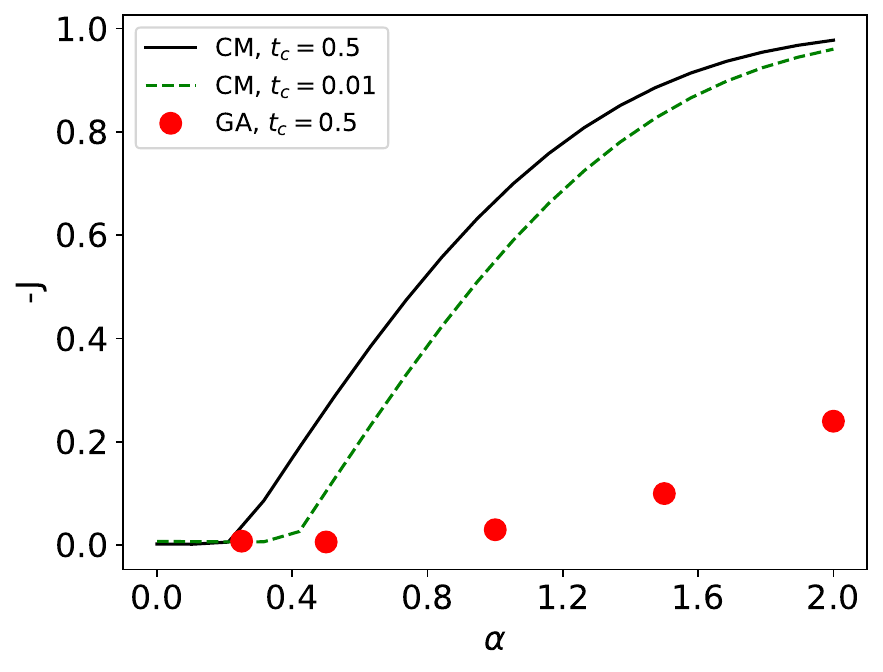}\\
 \textbf{(b)}\\
 \includegraphics[width=\columnwidth]{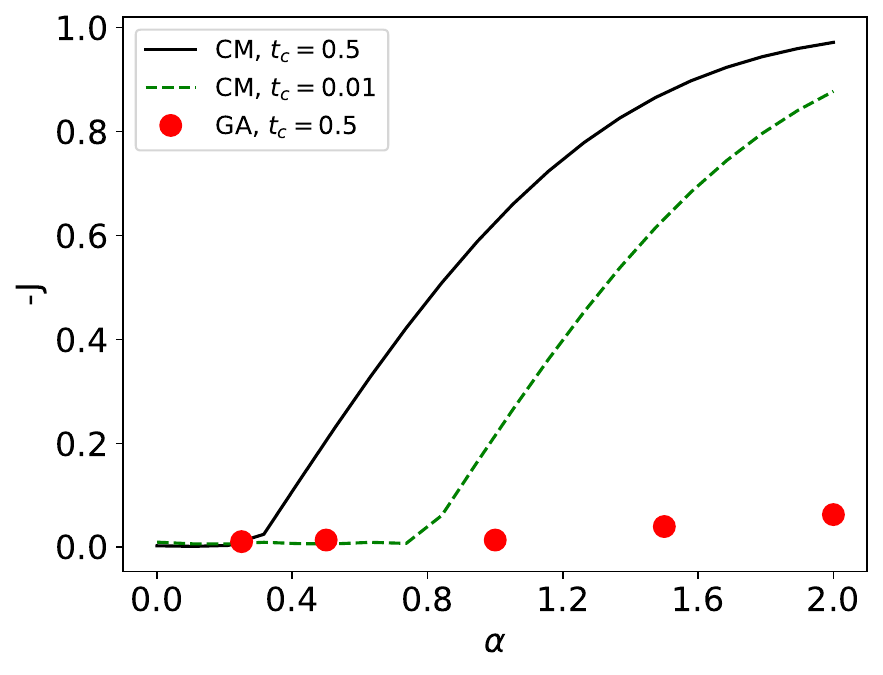}
\caption{Minimum trace distance between a coherent target state $|\alpha\rangle\langle\alpha|$ and the final state of the system obtained with a collision model-like process with optimal $\chi$ (lines) and with the fully optimized ancilla state sequence (dots), for $T=5$ \textbf{(a)} and for $T=10$ \textbf{(b)}. Each point in the lines represents the minimum trace distance obtained within the acceptable range of $\chi$ (see Fig.~\ref{Figure:coherent1}).}\label{Figure:coherent2}
\end{figure}

To the best of our knowledge, no scheme has been proposed that is able to exactly prepare the coherent states in Eq.~\eqref{equation:coherentstates}, with qubit ancillae, not even in the limit of infinite collisions. However, coherent thermal states
\begin{equation}\label{equation:coherentstates}
    \rho(\beta, \alpha) = D(\alpha) \frac{e^{-\beta H_S}}{\tr(e^{-\beta H_S})} D (\alpha)^\dagger,
\end{equation}
can be prepared in the long-time limit if all the ancillae are initialized as 
\begin{equation}\label{equation:ancillacoherences}
    \rho_A = \frac{e^{-\beta H_A}}{\tr(e^{-\beta H_A})} -\chi\sigma_Y,
\end{equation}
where $\sigma_Y$ is the $Y$-Pauli matrix and $\chi \approx \frac{\alpha  g t_c}{4}$. Notice that not all values of $\chi$ are admissible, as $\rho_A$ needs to be positive semi-definite in order to be a density matrix. This puts a restriction on the values of $\alpha$ achievable within a bounded range values of $g$ and $t_c$, if we fix the value of $\beta$. For our purposes, we require $\beta >> 1$, so that the $\rho(\beta, \alpha)$ does not differ much from $\rho^{target}$. Fixing $\beta=5$, the acceptable values of $\chi$ lay in the range $\approx[-0.08,0.08]$ (for simplicity, we restrict to only real values of $\alpha$). Within this range, we can search for the optimal value $\bar{\chi}$ that minimize the trace distance between the state of the system after a time $T$ and the target state (cf. Fig.~\ref{Figure:coherent1}). We can then compare these results with our approach based on the genetic algorithm proposed in Section~\ref{section:geneticalgorithms:fixedlength}. We found that, again, an optimized ancilla state sequence allowed us to significantly speed up the process of state preparation, as shown in Fig.~\ref{Figure:coherent2} for different values of $\alpha$.

In contrast to preparation of thermal states, this time coherences in the ancillae are required for the preparation of the target. For instance, for $\alpha=1$, $t_c=0.5$, $n=10$ using diagonal ancilla state sequence, we found $J (\rho^n_S, \rho^{target})\approx-0.69$, in contrast to $J (\rho^n_S, \rho^{target})\approx-0.033$, obtained with generic ancilla states. Furthermore, the performance can be improved further by using the algorithm described in Section~\ref{section:geneticalgorithms:variablelength} for $T=5$, and including the optimization of $\omega_A$ and $g$. In this case, we reached $J (\rho^{n=25}_S, \rho^{target})\approx-0.015$ for $t_c=0.2$, $\omega_A\approx3.3$, $g\approx0.83$. Notice that the optimal coupling is slightly weaker and that the cavity and the ancillae are significantly detuned. Finally, using the same approach but with three level system ancillae instead of qubits, the performance was further improved, leading to $J (\rho^{n}_S, \rho^{target})\approx-0.009$.

\subsection{Squeezed states}\label{section:results:squeezedstates}
Another class of important cavity states is that of squeezed states~\cite{loudon2000quantum, Walls1983}. Let us consider the preparation of squeezed vacuum states
\begin{equation}\label{equation:squeezedstates}
    \rho^{target} = S(\zeta)|0\rangle \langle 0| S (\zeta)^\dagger,
\end{equation}
where $S(\zeta)=\exp[\frac{1}{2}(\zeta^* a^2 - \zeta {a^\dagger}^2)]$ and $\zeta\in\mathbb{C}$, in general. 

We started by applying the algorithm described in Section~\ref{section:geneticalgorithms:variablelength} to prepare a squeezed state with $\zeta = 0.5$ within a time $T=5$. This time, using a linear system-ancillae interaction with qubit ancillae was not sufficient to prepare a target state. This did not change even when we replaced the qubits with three level systems (in both cases $J (\rho^n_S, \rho^{target})\approx -0.34$). On the other end, when we included the nonlinear interaction described in in Eq.~\eqref{equation:nonlinearIqubit}, we were able to successfully prepare the desired target state, even with qubit ancillae ($J (\rho^n_S, \rho^{target})\approx -0.0083$). This highlights how a linear interaction is not sufficient to allow for the preparation of squeezed states, even whith engineered ancilla states. The optimum is achieved after $n=79$ collisions ($t_c\approx0.06$), no linear contribution to the interaction ($g_{l}\approx10^{-5}$ and $g_{nl}\approx-0.45$) and detuned ancillae ($\omega_A \approx 1.81 \omega_S$). While we did not find any collision model with single qubit ancillae that leads to a stationary squeezed state, a collision model and a corresponding Master Equation has been derived for entangled qubit ancillae interacting linearly with the cavity in Ref.~\cite{Miao_2017}. From our numerical simulations, however, convergence to the target state through such master equation is significantly slower: $-J (\rho(T=5), \rho^{target})> 0.1 $ for $t_c\in[0.01,0.5]$.

We also considered the preparation of a highly squeezed state, with $\zeta=1$ within the same time $T=5$. Using qubit ancillae, we obtained $J (\rho^n_S, \rho^{target})\approx -0.057$. Aside from the ancilla states, optimal parameters do not differ much from the $\zeta=0.5$ case: $n=73$ ($t_c\approx0.07$), $g_{l}\approx10^{-5}$ and $g_{nl}\approx-0.53$, $\omega_A \approx 1.81 \omega_S$. There does not seem to be any substantial improvement when using three level systems as ancillae instead of qubits [$J (\rho^n_S, \rho^{target})\approx -0.056$]. 

\subsection{Non-Gaussian states}\label{section:results:nongaussianstates}
Although all the cavity states considered above are Gaussian, the methodology presented here is not limited to the preparation of this class of states. This is very important as the ability to prepare non-Gaussian states is crucial, particularly in the context of continuous-variable quantum information and computing~\cite{PhysRevLett.89.137903, PhysRevLett.102.120501, Ferraro_2005}. In order to illustrate the capabilities of our approach for the engineering of non-Gaussian states we mention that a Fock state with a single cavity photon can be prepared in the linear regime by appropriately selecting $T$ and $g$ and letting the system interact with a single ancilla, prepared in the excited state. We checked that the algorithm in Section~\ref{section:geneticalgorithms:variablelength} can prepare such state, using exactly the strategy we just described, obtaining $J (\rho^{i=1}_S, |1\rangle \langle 1|)\approx - 10^{-4}$. It is then interesting to determine the maximum degree of non-Gaussianity that can be achieved using our approach.

A  way of quantifying the degree of non-Gaussianity of a quantum state is to estimate the {\it similarity} of the state under scrutiny from the Gaussian state $\rho_G$ with the same mean and variance. In Ref.~\cite{PhysRevA.82.052341}, it is proposed to use the quantum relative entropy to quantify such similarity, thus introducing the quantity
\begin{equation}\label{equation:relativentropy}
    S(\rho||\rho_G) = \tr(\rho\log\rho-\rho\log\rho_G),
\end{equation}
which can be computed as
\begin{equation}
    S(\rho||\rho_G) = h(\sqrt{\det\sigma}) - S(\rho),
\end{equation}
where $h(x) = (x + \frac{1}{2})\log(x + \frac{1}{2}) - (x - \frac{1}{2})\log(x - \frac{1}{2})$, $S(\rho) = -\rho\log\rho$ is the von Neumann entropy of $\rho$ and $\sigma$ is the associated covariance matrix 
\begin{equation}
    \sigma_{ij} = \langle X_{\phi_i} X_{\phi_j} + X_{\phi_j} X_{\phi_i}\rangle - \langle X_{\phi_i} \rangle \langle X_{\phi_j}\rangle,
\end{equation}
for $\phi_{i,j}\in\{0,\pi/2\}$ and where we have defined the field quadratures 
\begin{equation}
    X_\phi=\frac{ae^{-i\phi}+a^\dagger e^{i\phi}}{2}.
\end{equation}
Our strategy was  to use $S(\rho^n_S\|\rho_G)$ as a \textit{fitness} function for the algorithm in Sec.~\ref{section:geneticalgorithms:variablelength} instead of $J (\rho^n_S, \rho^{target})$ with a specific target, in order to obtain a state that is as non-Gaussian as possibly achievable. In our study, we considered only qubit ancillae and found $S(\rho^n_S\|\rho_G)\approx1.43$ and $S(\rho^n_S\|\rho_G)\approx2.42$ for linear ($H_{nl} = 0$) and nonlinear [$H_{nl}$ as in Eq.~\eqref{equation:nonlinearIqubit}] couplings, respectively.  While, for reference, we have $S(|k\rangle \langle k|\|\rho_G)\approx1.38, 1.91, 2.25, 2.50$, for Fock states $|k\rangle$ with $k=1, 2, 3, 4$ photons respectively, the final states do not resemble any known class of non-Gaussian state.

\section{Conclusions}\label{section:conclusions}
We have achieved fast thermalization and state preparation of the field of a single-mode cavity  using engineered collision-like processes, where a stream of suitably prepared ancillae sequentially interact n with the system. We optimized the ancilla state sequence, along with some constant parameters of the dynamics and minimized the trace distance with the desired target state, using genetic algorithms. We considered both two- and three-level system ancillae. We were able to significantly accelerate thermalization and preparation of coherent states and to effectively prepare squeezed cavity states. We demonstrated that the optimal ancilla sequences and collision times are non-trivial and that coherences in the ancillae are not required to accelerate thermalization, both in the case of heating and cooling. We also verified that a non-linear interaction is a requirement for squeezed state preparation, but not for the other states considered. We prepared highly non-Gaussian states by deploying the same genetic approach to maximize the relative entropy between the final state and the corresponding Gaussian state.

Our work contributes to the development of strategies for state preparation and reservoir engineering by offering a compromise between fast time-dependent Hamiltonian control and steady-state environment engineering. Moreover, it further improves our understanding of the resources needed to realize and speed-up the preparation of specific cavity states. 
An interesting venue for future work is the development of analytical strategies to predict which states can be prepared with a given set of resources ahead of the numerical simulations. It would also be interesting to study the impact of using a feedback-control loop to determine the state of the next ancilla at each step. This would require, however, a different approach, based, for example, on techniques borrowed from reinforcement learning.

\section*{Data availability statement}
The data that support the findings of this study and the code implementing the algorithms presented in Sec.~\ref{section:geneticalgorithms} are openly available at the following \href{https://github.com/SofiaSgroi/Speeding-up-thermalization-and-quantum-state-preparation-through-engineered-quantum-collisions.git}{github repository}.

\acknowledgements
We acknowledge support from the  European Union - NextGenerationEU through the Italian Ministry of University and Research (MUR) under PNRR - M4C2-I1.3 Project
PE-00000019 ``HEAL ITALIA" (CUP B73C22001250006
), the ``Italian National Quantum Science and Technology Institute (NQSTI)" (PE0000023) - SPOKE 2 through project ASpEQCt, the ``National Centre for HPC, Big Data and Quantum Computing (HPC)" (CN00000013) - SPOKE 10 through project HyQELM, the Italian MUR under PRIN Project ``Quantum Reservoir Computing (QuReCo)" (No. 2022FEXLYB). MP acknowledges the European Union's Horizon Europe EIC-Pathfinder project QuCoM (101046973), the UK EPSRC (grants EP/T028424/1 and EP/X021505/1), the Department for the Economy of Northern Ireland under the US-Ireland R\&D Partnership Programme.

\bibliography{biblio}

\begin{thebibliography}{53}%
\makeatletter
\providecommand \@ifxundefined [1]{%
 \@ifx{#1\undefined}
}%
\providecommand \@ifnum [1]{%
 \ifnum #1\expandafter \@firstoftwo
 \else \expandafter \@secondoftwo
 \fi
}%
\providecommand \@ifx [1]{%
 \ifx #1\expandafter \@firstoftwo
 \else \expandafter \@secondoftwo
 \fi
}%
\providecommand \natexlab [1]{#1}%
\providecommand \enquote  [1]{``#1''}%
\providecommand \bibnamefont  [1]{#1}%
\providecommand \bibfnamefont [1]{#1}%
\providecommand \citenamefont [1]{#1}%
\providecommand \href@noop [0]{\@secondoftwo}%
\providecommand \href [0]{\begingroup \@sanitize@url \@href}%
\providecommand \@href[1]{\@@startlink{#1}\@@href}%
\providecommand \@@href[1]{\endgroup#1\@@endlink}%
\providecommand \@sanitize@url [0]{\catcode `\\12\catcode `\$12\catcode `\&12\catcode `\#12\catcode `\^12\catcode `\_12\catcode `\%12\relax}%
\providecommand \@@startlink[1]{}%
\providecommand \@@endlink[0]{}%
\providecommand \url  [0]{\begingroup\@sanitize@url \@url }%
\providecommand \@url [1]{\endgroup\@href {#1}{\urlprefix }}%
\providecommand \urlprefix  [0]{URL }%
\providecommand \Eprint [0]{\href }%
\providecommand \doibase [0]{https://doi.org/}%
\providecommand \selectlanguage [0]{\@gobble}%
\providecommand \bibinfo  [0]{\@secondoftwo}%
\providecommand \bibfield  [0]{\@secondoftwo}%
\providecommand \translation [1]{[#1]}%
\providecommand \BibitemOpen [0]{}%
\providecommand \bibitemStop [0]{}%
\providecommand \bibitemNoStop [0]{.\EOS\space}%
\providecommand \EOS [0]{\spacefactor3000\relax}%
\providecommand \BibitemShut  [1]{\csname bibitem#1\endcsname}%
\let\auto@bib@innerbib\@empty
\bibitem [{\citenamefont {Plesch}\ and\ \citenamefont {Brukner}(2011)}]{PhysRevA.83.032302}%
  \BibitemOpen
  \bibfield  {author} {\bibinfo {author} {\bibfnamefont {M.}~\bibnamefont {Plesch}}\ and\ \bibinfo {author} {\bibfnamefont {{\v C}.}~\bibnamefont {Brukner}},\ }\bibfield  {title} {\bibinfo {title} {Quantum-state preparation with universal gate decompositions},\ }\href {https://doi.org/10.1103/PhysRevA.83.032302} {\bibfield  {journal} {\bibinfo  {journal} {Phys. Rev. A}\ }\textbf {\bibinfo {volume} {83}},\ \bibinfo {pages} {032302} (\bibinfo {year} {2011})}\BibitemShut {NoStop}%
\bibitem [{\citenamefont {Berry}(2009)}]{Berry_2009}%
  \BibitemOpen
  \bibfield  {author} {\bibinfo {author} {\bibfnamefont {M.~V.}\ \bibnamefont {Berry}},\ }\bibfield  {title} {\bibinfo {title} {Transitionless quantum driving},\ }\href {https://doi.org/10.1088/1751-8113/42/36/365303} {\bibfield  {journal} {\bibinfo  {journal} {Journal of Physics A: Mathematical and Theoretical}\ }\textbf {\bibinfo {volume} {42}},\ \bibinfo {pages} {365303} (\bibinfo {year} {2009})}\BibitemShut {NoStop}%
\bibitem [{\citenamefont {Gu\'ery-Odelin}\ \emph {et~al.}(2019)\citenamefont {Gu\'ery-Odelin}, \citenamefont {Ruschhaupt}, \citenamefont {Kiely}, \citenamefont {Torrontegui}, \citenamefont {Mart\'{\i}nez-Garaot},\ and\ \citenamefont {Muga}}]{RevModPhys.91.045001}%
  \BibitemOpen
  \bibfield  {author} {\bibinfo {author} {\bibfnamefont {D.}~\bibnamefont {Gu\'ery-Odelin}}, \bibinfo {author} {\bibfnamefont {A.}~\bibnamefont {Ruschhaupt}}, \bibinfo {author} {\bibfnamefont {A.}~\bibnamefont {Kiely}}, \bibinfo {author} {\bibfnamefont {E.}~\bibnamefont {Torrontegui}}, \bibinfo {author} {\bibfnamefont {S.}~\bibnamefont {Mart\'{\i}nez-Garaot}},\ and\ \bibinfo {author} {\bibfnamefont {J.~G.}\ \bibnamefont {Muga}},\ }\bibfield  {title} {\bibinfo {title} {Shortcuts to adiabaticity: Concepts, methods, and applications},\ }\href {https://doi.org/10.1103/RevModPhys.91.045001} {\bibfield  {journal} {\bibinfo  {journal} {Rev. Mod. Phys.}\ }\textbf {\bibinfo {volume} {91}},\ \bibinfo {pages} {045001} (\bibinfo {year} {2019})}\BibitemShut {NoStop}%
\bibitem [{\citenamefont {Glaser}\ \emph {et~al.}(2015)\citenamefont {Glaser}, \citenamefont {Boscain}, \citenamefont {Calarco}, \citenamefont {Koch}, \citenamefont {K{\"o}ckenberger}, \citenamefont {Kosloff}, \citenamefont {Kuprov}, \citenamefont {Luy}, \citenamefont {Schirmer}, \citenamefont {Schulte-Herbr{\"u}ggen}, \citenamefont {Sugny},\ and\ \citenamefont {Wilhelm}}]{Glaser2015}%
  \BibitemOpen
  \bibfield  {author} {\bibinfo {author} {\bibfnamefont {S.~J.}\ \bibnamefont {Glaser}}, \bibinfo {author} {\bibfnamefont {U.}~\bibnamefont {Boscain}}, \bibinfo {author} {\bibfnamefont {T.}~\bibnamefont {Calarco}}, \bibinfo {author} {\bibfnamefont {C.~P.}\ \bibnamefont {Koch}}, \bibinfo {author} {\bibfnamefont {W.}~\bibnamefont {K{\"o}ckenberger}}, \bibinfo {author} {\bibfnamefont {R.}~\bibnamefont {Kosloff}}, \bibinfo {author} {\bibfnamefont {I.}~\bibnamefont {Kuprov}}, \bibinfo {author} {\bibfnamefont {B.}~\bibnamefont {Luy}}, \bibinfo {author} {\bibfnamefont {S.}~\bibnamefont {Schirmer}}, \bibinfo {author} {\bibfnamefont {T.}~\bibnamefont {Schulte-Herbr{\"u}ggen}}, \bibinfo {author} {\bibfnamefont {D.}~\bibnamefont {Sugny}},\ and\ \bibinfo {author} {\bibfnamefont {F.~K.}\ \bibnamefont {Wilhelm}},\ }\bibfield  {title} {\bibinfo {title} {Training schr{\"o}dinger's cat: quantum optimal control},\ }\href {https://doi.org/10.1140/epjd/e2015-60464-1} {\bibfield  {journal} {\bibinfo  {journal} {The European
  Physical Journal D}\ }\textbf {\bibinfo {volume} {69}},\ \bibinfo {pages} {279} (\bibinfo {year} {2015})}\BibitemShut {NoStop}%
\bibitem [{\citenamefont {Boscain}\ \emph {et~al.}(2021)\citenamefont {Boscain}, \citenamefont {Sigalotti},\ and\ \citenamefont {Sugny}}]{PRXQuantum.2.030203}%
  \BibitemOpen
  \bibfield  {author} {\bibinfo {author} {\bibfnamefont {U.}~\bibnamefont {Boscain}}, \bibinfo {author} {\bibfnamefont {M.}~\bibnamefont {Sigalotti}},\ and\ \bibinfo {author} {\bibfnamefont {D.}~\bibnamefont {Sugny}},\ }\bibfield  {title} {\bibinfo {title} {Introduction to the pontryagin maximum principle for quantum optimal control},\ }\href {https://doi.org/10.1103/PRXQuantum.2.030203} {\bibfield  {journal} {\bibinfo  {journal} {PRX Quantum}\ }\textbf {\bibinfo {volume} {2}},\ \bibinfo {pages} {030203} (\bibinfo {year} {2021})}\BibitemShut {NoStop}%
\bibitem [{\citenamefont {Wilhelm}\ \emph {et~al.}(2020)\citenamefont {Wilhelm}, \citenamefont {Kirchhoff}, \citenamefont {Machnes}, \citenamefont {Wittler},\ and\ \citenamefont {Sugny}}]{wilhelm2020introductionoptimalcontrolquantum}%
  \BibitemOpen
  \bibfield  {author} {\bibinfo {author} {\bibfnamefont {F.~K.}\ \bibnamefont {Wilhelm}}, \bibinfo {author} {\bibfnamefont {S.}~\bibnamefont {Kirchhoff}}, \bibinfo {author} {\bibfnamefont {S.}~\bibnamefont {Machnes}}, \bibinfo {author} {\bibfnamefont {N.}~\bibnamefont {Wittler}},\ and\ \bibinfo {author} {\bibfnamefont {D.}~\bibnamefont {Sugny}},\ }\href {https://arxiv.org/abs/2003.10132} {\bibinfo {title} {An introduction into optimal control for quantum technologies}} (\bibinfo {year} {2020}),\ \Eprint {https://arxiv.org/abs/2003.10132} {arXiv:2003.10132 [quant-ph]} \BibitemShut {NoStop}%
\bibitem [{\citenamefont {Bukov}\ \emph {et~al.}(2018)\citenamefont {Bukov}, \citenamefont {Day}, \citenamefont {Sels}, \citenamefont {Weinberg}, \citenamefont {Polkovnikov},\ and\ \citenamefont {Mehta}}]{PhysRevX.8.031086}%
  \BibitemOpen
  \bibfield  {author} {\bibinfo {author} {\bibfnamefont {M.}~\bibnamefont {Bukov}}, \bibinfo {author} {\bibfnamefont {A.~G.~R.}\ \bibnamefont {Day}}, \bibinfo {author} {\bibfnamefont {D.}~\bibnamefont {Sels}}, \bibinfo {author} {\bibfnamefont {P.}~\bibnamefont {Weinberg}}, \bibinfo {author} {\bibfnamefont {A.}~\bibnamefont {Polkovnikov}},\ and\ \bibinfo {author} {\bibfnamefont {P.}~\bibnamefont {Mehta}},\ }\bibfield  {title} {\bibinfo {title} {Reinforcement learning in different phases of quantum control},\ }\href {https://doi.org/10.1103/PhysRevX.8.031086} {\bibfield  {journal} {\bibinfo  {journal} {Phys. Rev. X}\ }\textbf {\bibinfo {volume} {8}},\ \bibinfo {pages} {031086} (\bibinfo {year} {2018})}\BibitemShut {NoStop}%
\bibitem [{\citenamefont {Zhang}\ \emph {et~al.}(2019)\citenamefont {Zhang}, \citenamefont {Wei}, \citenamefont {Asad}, \citenamefont {Yang},\ and\ \citenamefont {Wang}}]{Zhang2019}%
  \BibitemOpen
  \bibfield  {author} {\bibinfo {author} {\bibfnamefont {X.-M.}\ \bibnamefont {Zhang}}, \bibinfo {author} {\bibfnamefont {Z.}~\bibnamefont {Wei}}, \bibinfo {author} {\bibfnamefont {R.}~\bibnamefont {Asad}}, \bibinfo {author} {\bibfnamefont {X.-C.}\ \bibnamefont {Yang}},\ and\ \bibinfo {author} {\bibfnamefont {X.}~\bibnamefont {Wang}},\ }\bibfield  {title} {\bibinfo {title} {When does reinforcement learning stand out in quantum control? a comparative study on state preparation},\ }\href {https://doi.org/10.1038/s41534-019-0201-8} {\bibfield  {journal} {\bibinfo  {journal} {npj Quantum Information}\ }\textbf {\bibinfo {volume} {5}},\ \bibinfo {pages} {85} (\bibinfo {year} {2019})}\BibitemShut {NoStop}%
\bibitem [{\citenamefont {Brown}\ \emph {et~al.}(2021)\citenamefont {Brown}, \citenamefont {Sgroi}, \citenamefont {Giannelli}, \citenamefont {Paraoanu}, \citenamefont {Paladino}, \citenamefont {Falci}, \citenamefont {Paternostro},\ and\ \citenamefont {Ferraro}}]{Brown2021}%
  \BibitemOpen
  \bibfield  {author} {\bibinfo {author} {\bibfnamefont {J.}~\bibnamefont {Brown}}, \bibinfo {author} {\bibfnamefont {S.}~\bibnamefont {Sgroi}}, \bibinfo {author} {\bibfnamefont {L.}~\bibnamefont {Giannelli}}, \bibinfo {author} {\bibfnamefont {G.~S.}\ \bibnamefont {Paraoanu}}, \bibinfo {author} {\bibfnamefont {E.}~\bibnamefont {Paladino}}, \bibinfo {author} {\bibfnamefont {G.}~\bibnamefont {Falci}}, \bibinfo {author} {\bibfnamefont {M.}~\bibnamefont {Paternostro}},\ and\ \bibinfo {author} {\bibfnamefont {A.}~\bibnamefont {Ferraro}},\ }\bibfield  {title} {\bibinfo {title} {Reinforcement learning-enhanced protocols for coherent population-transfer in three-level quantum systems},\ }\href {https://iopscience.iop.org/article/10.1088/1367-2630/ac2393} {\bibfield  {journal} {\bibinfo  {journal} {New Journal of Physics}\ }\textbf {\bibinfo {volume} {23}},\ \bibinfo {pages} {093035} (\bibinfo {year} {2021})}\BibitemShut {NoStop}%
\bibitem [{\citenamefont {Kraus}\ \emph {et~al.}(2008)\citenamefont {Kraus}, \citenamefont {B\"uchler}, \citenamefont {Diehl}, \citenamefont {Kantian}, \citenamefont {Micheli},\ and\ \citenamefont {Zoller}}]{PhysRevA.78.042307}%
  \BibitemOpen
  \bibfield  {author} {\bibinfo {author} {\bibfnamefont {B.}~\bibnamefont {Kraus}}, \bibinfo {author} {\bibfnamefont {H.~P.}\ \bibnamefont {B\"uchler}}, \bibinfo {author} {\bibfnamefont {S.}~\bibnamefont {Diehl}}, \bibinfo {author} {\bibfnamefont {A.}~\bibnamefont {Kantian}}, \bibinfo {author} {\bibfnamefont {A.}~\bibnamefont {Micheli}},\ and\ \bibinfo {author} {\bibfnamefont {P.}~\bibnamefont {Zoller}},\ }\bibfield  {title} {\bibinfo {title} {Preparation of entangled states by quantum markov processes},\ }\href {https://doi.org/10.1103/PhysRevA.78.042307} {\bibfield  {journal} {\bibinfo  {journal} {Phys. Rev. A}\ }\textbf {\bibinfo {volume} {78}},\ \bibinfo {pages} {042307} (\bibinfo {year} {2008})}\BibitemShut {NoStop}%
\bibitem [{\citenamefont {Carvalho}\ \emph {et~al.}(2001)\citenamefont {Carvalho}, \citenamefont {Milman}, \citenamefont {de~Matos~Filho},\ and\ \citenamefont {Davidovich}}]{PhysRevLett.86.4988}%
  \BibitemOpen
  \bibfield  {author} {\bibinfo {author} {\bibfnamefont {A.~R.~R.}\ \bibnamefont {Carvalho}}, \bibinfo {author} {\bibfnamefont {P.}~\bibnamefont {Milman}}, \bibinfo {author} {\bibfnamefont {R.~L.}\ \bibnamefont {de~Matos~Filho}},\ and\ \bibinfo {author} {\bibfnamefont {L.}~\bibnamefont {Davidovich}},\ }\bibfield  {title} {\bibinfo {title} {Decoherence, pointer engineering, and quantum state protection},\ }\href {https://doi.org/10.1103/PhysRevLett.86.4988} {\bibfield  {journal} {\bibinfo  {journal} {Phys. Rev. Lett.}\ }\textbf {\bibinfo {volume} {86}},\ \bibinfo {pages} {4988} (\bibinfo {year} {2001})}\BibitemShut {NoStop}%
\bibitem [{\citenamefont {Sarlette}\ \emph {et~al.}(2011)\citenamefont {Sarlette}, \citenamefont {Raimond}, \citenamefont {Brune},\ and\ \citenamefont {Rouchon}}]{PhysRevLett.107.010402}%
  \BibitemOpen
  \bibfield  {author} {\bibinfo {author} {\bibfnamefont {A.}~\bibnamefont {Sarlette}}, \bibinfo {author} {\bibfnamefont {J.~M.}\ \bibnamefont {Raimond}}, \bibinfo {author} {\bibfnamefont {M.}~\bibnamefont {Brune}},\ and\ \bibinfo {author} {\bibfnamefont {P.}~\bibnamefont {Rouchon}},\ }\bibfield  {title} {\bibinfo {title} {Stabilization of nonclassical states of the radiation field in a cavity by reservoir engineering},\ }\href {https://doi.org/10.1103/PhysRevLett.107.010402} {\bibfield  {journal} {\bibinfo  {journal} {Phys. Rev. Lett.}\ }\textbf {\bibinfo {volume} {107}},\ \bibinfo {pages} {010402} (\bibinfo {year} {2011})}\BibitemShut {NoStop}%
\bibitem [{\citenamefont {Sarlette}\ \emph {et~al.}(2012)\citenamefont {Sarlette}, \citenamefont {Leghtas}, \citenamefont {Brune}, \citenamefont {Raimond},\ and\ \citenamefont {Rouchon}}]{PhysRevA.86.012114}%
  \BibitemOpen
  \bibfield  {author} {\bibinfo {author} {\bibfnamefont {A.}~\bibnamefont {Sarlette}}, \bibinfo {author} {\bibfnamefont {Z.}~\bibnamefont {Leghtas}}, \bibinfo {author} {\bibfnamefont {M.}~\bibnamefont {Brune}}, \bibinfo {author} {\bibfnamefont {J.~M.}\ \bibnamefont {Raimond}},\ and\ \bibinfo {author} {\bibfnamefont {P.}~\bibnamefont {Rouchon}},\ }\bibfield  {title} {\bibinfo {title} {Stabilization of nonclassical states of one- and two-mode radiation fields by reservoir engineering},\ }\href {https://doi.org/10.1103/PhysRevA.86.012114} {\bibfield  {journal} {\bibinfo  {journal} {Phys. Rev. A}\ }\textbf {\bibinfo {volume} {86}},\ \bibinfo {pages} {012114} (\bibinfo {year} {2012})}\BibitemShut {NoStop}%
\bibitem [{\citenamefont {Mukherjee}\ \emph {et~al.}(2013)\citenamefont {Mukherjee}, \citenamefont {Carlini}, \citenamefont {Mari}, \citenamefont {Caneva}, \citenamefont {Montangero}, \citenamefont {Calarco}, \citenamefont {Fazio},\ and\ \citenamefont {Giovannetti}}]{PhysRevA.88.062326}%
  \BibitemOpen
  \bibfield  {author} {\bibinfo {author} {\bibfnamefont {V.}~\bibnamefont {Mukherjee}}, \bibinfo {author} {\bibfnamefont {A.}~\bibnamefont {Carlini}}, \bibinfo {author} {\bibfnamefont {A.}~\bibnamefont {Mari}}, \bibinfo {author} {\bibfnamefont {T.}~\bibnamefont {Caneva}}, \bibinfo {author} {\bibfnamefont {S.}~\bibnamefont {Montangero}}, \bibinfo {author} {\bibfnamefont {T.}~\bibnamefont {Calarco}}, \bibinfo {author} {\bibfnamefont {R.}~\bibnamefont {Fazio}},\ and\ \bibinfo {author} {\bibfnamefont {V.}~\bibnamefont {Giovannetti}},\ }\bibfield  {title} {\bibinfo {title} {Speeding up and slowing down the relaxation of a qubit by optimal control},\ }\href {https://doi.org/10.1103/PhysRevA.88.062326} {\bibfield  {journal} {\bibinfo  {journal} {Phys. Rev. A}\ }\textbf {\bibinfo {volume} {88}},\ \bibinfo {pages} {062326} (\bibinfo {year} {2013})}\BibitemShut {NoStop}%
\bibitem [{\citenamefont {Dann}\ \emph {et~al.}(2019)\citenamefont {Dann}, \citenamefont {Tobalina},\ and\ \citenamefont {Kosloff}}]{PhysRevLett.122.250402}%
  \BibitemOpen
  \bibfield  {author} {\bibinfo {author} {\bibfnamefont {R.}~\bibnamefont {Dann}}, \bibinfo {author} {\bibfnamefont {A.}~\bibnamefont {Tobalina}},\ and\ \bibinfo {author} {\bibfnamefont {R.}~\bibnamefont {Kosloff}},\ }\bibfield  {title} {\bibinfo {title} {Shortcut to equilibration of an open quantum system},\ }\href {https://doi.org/10.1103/PhysRevLett.122.250402} {\bibfield  {journal} {\bibinfo  {journal} {Phys. Rev. Lett.}\ }\textbf {\bibinfo {volume} {122}},\ \bibinfo {pages} {250402} (\bibinfo {year} {2019})}\BibitemShut {NoStop}%
\bibitem [{\citenamefont {Dann}\ \emph {et~al.}(2020)\citenamefont {Dann}, \citenamefont {Tobalina},\ and\ \citenamefont {Kosloff}}]{PhysRevA.101.052102}%
  \BibitemOpen
  \bibfield  {author} {\bibinfo {author} {\bibfnamefont {R.}~\bibnamefont {Dann}}, \bibinfo {author} {\bibfnamefont {A.}~\bibnamefont {Tobalina}},\ and\ \bibinfo {author} {\bibfnamefont {R.}~\bibnamefont {Kosloff}},\ }\bibfield  {title} {\bibinfo {title} {Fast route to equilibration},\ }\href {https://doi.org/10.1103/PhysRevA.101.052102} {\bibfield  {journal} {\bibinfo  {journal} {Phys. Rev. A}\ }\textbf {\bibinfo {volume} {101}},\ \bibinfo {pages} {052102} (\bibinfo {year} {2020})}\BibitemShut {NoStop}%
\bibitem [{\citenamefont {Kallush}\ \emph {et~al.}(2022)\citenamefont {Kallush}, \citenamefont {Dann},\ and\ \citenamefont {Kosloff}}]{Kallush_2022}%
  \BibitemOpen
  \bibfield  {author} {\bibinfo {author} {\bibfnamefont {S.}~\bibnamefont {Kallush}}, \bibinfo {author} {\bibfnamefont {R.}~\bibnamefont {Dann}},\ and\ \bibinfo {author} {\bibfnamefont {R.}~\bibnamefont {Kosloff}},\ }\bibfield  {title} {\bibinfo {title} {Controlling the uncontrollable: Quantum control of open-system dynamics},\ }\href {https://doi.org/10.1126/sciadv.add0828} {\bibfield  {journal} {\bibinfo  {journal} {Science Advances}\ }\textbf {\bibinfo {volume} {8}},\ \bibinfo {pages} {eadd0828} (\bibinfo {year} {2022})},\ \Eprint {https://arxiv.org/abs/https://www.science.org/doi/pdf/10.1126/sciadv.add0828} {https://www.science.org/doi/pdf/10.1126/sciadv.add0828} \BibitemShut {NoStop}%
\bibitem [{\citenamefont {Pedram}\ \emph {et~al.}(2023)\citenamefont {Pedram}, \citenamefont {Kad{\i}o{\u g}lu}, \citenamefont {Kabak{\c c}{\i}o{\u g}lu},\ and\ \citenamefont {M{\"u}stecapl{\i}o{\u g}lu}}]{Pedram_2023}%
  \BibitemOpen
  \bibfield  {author} {\bibinfo {author} {\bibfnamefont {A.}~\bibnamefont {Pedram}}, \bibinfo {author} {\bibfnamefont {S.~C.}\ \bibnamefont {Kad{\i}o{\u g}lu}}, \bibinfo {author} {\bibfnamefont {A.}~\bibnamefont {Kabak{\c c}{\i}o{\u g}lu}},\ and\ \bibinfo {author} {\bibfnamefont {{\"O}.~E.}\ \bibnamefont {M{\"u}stecapl{\i}o{\u g}lu}},\ }\bibfield  {title} {\bibinfo {title} {A quantum {O}tto engine with shortcuts to thermalization and adiabaticity},\ }\href {https://doi.org/10.1088/1367-2630/ad0857} {\bibfield  {journal} {\bibinfo  {journal} {New Journal of Physics}\ }\textbf {\bibinfo {volume} {25}},\ \bibinfo {pages} {113014} (\bibinfo {year} {2023})}\BibitemShut {NoStop}%
\bibitem [{\citenamefont {Ciccarello}\ \emph {et~al.}(2022)\citenamefont {Ciccarello}, \citenamefont {Lorenzo}, \citenamefont {Giovannetti},\ and\ \citenamefont {Palma}}]{CICCARELLO20221}%
  \BibitemOpen
  \bibfield  {author} {\bibinfo {author} {\bibfnamefont {F.}~\bibnamefont {Ciccarello}}, \bibinfo {author} {\bibfnamefont {S.}~\bibnamefont {Lorenzo}}, \bibinfo {author} {\bibfnamefont {V.}~\bibnamefont {Giovannetti}},\ and\ \bibinfo {author} {\bibfnamefont {G.~M.}\ \bibnamefont {Palma}},\ }\bibfield  {title} {\bibinfo {title} {Quantum collision models: Open system dynamics from repeated interactions},\ }\href {https://doi.org/https://doi.org/10.1016/j.physrep.2022.01.001} {\bibfield  {journal} {\bibinfo  {journal} {Physics Reports}\ }\textbf {\bibinfo {volume} {954}},\ \bibinfo {pages} {1} (\bibinfo {year} {2022})}\BibitemShut {NoStop}%
\bibitem [{\citenamefont {Ciccarello}(2017)}]{ciccarello2017collision}%
  \BibitemOpen
  \bibfield  {author} {\bibinfo {author} {\bibfnamefont {F.}~\bibnamefont {Ciccarello}},\ }\bibfield  {title} {\bibinfo {title} {Collision models in quantum optics},\ }\href {https://doi.org/10.1515/qmetro-2017-000} {\bibfield  {journal} {\bibinfo  {journal} {Quantum Measurements and Quantum Metrology}\ }\textbf {\bibinfo {volume} {4}},\ \bibinfo {pages} {53} (\bibinfo {year} {2017})}\BibitemShut {NoStop}%
\bibitem [{\citenamefont {Alicki}\ and\ \citenamefont {Lendi}(2007)}]{alicki2007quantum}%
  \BibitemOpen
  \bibfield  {author} {\bibinfo {author} {\bibfnamefont {R.}~\bibnamefont {Alicki}}\ and\ \bibinfo {author} {\bibfnamefont {K.}~\bibnamefont {Lendi}},\ }\href@noop {} {\emph {\bibinfo {title} {Quantum dynamical semigroups and applications}}},\ Vol.\ \bibinfo {volume} {717}\ (\bibinfo  {publisher} {Springer},\ \bibinfo {year} {2007})\BibitemShut {NoStop}%
\bibitem [{\citenamefont {Scarani}\ \emph {et~al.}(2002)\citenamefont {Scarani}, \citenamefont {Ziman}, \citenamefont {\ifmmode \check{S}\else \v{S}\fi{}telmachovi\ifmmode~\check{c}\else \v{c}\fi{}}, \citenamefont {Gisin},\ and\ \citenamefont {Bu\ifmmode~\check{z}\else \v{z}\fi{}ek}}]{PhysRevLett.88.097905}%
  \BibitemOpen
  \bibfield  {author} {\bibinfo {author} {\bibfnamefont {V.}~\bibnamefont {Scarani}}, \bibinfo {author} {\bibfnamefont {M.}~\bibnamefont {Ziman}}, \bibinfo {author} {\bibfnamefont {P.}~\bibnamefont {\ifmmode \check{S}\else \v{S}\fi{}telmachovi\ifmmode~\check{c}\else \v{c}\fi{}}}, \bibinfo {author} {\bibfnamefont {N.}~\bibnamefont {Gisin}},\ and\ \bibinfo {author} {\bibfnamefont {V.}~\bibnamefont {Bu\ifmmode~\check{z}\else \v{z}\fi{}ek}},\ }\bibfield  {title} {\bibinfo {title} {Thermalizing quantum machines: Dissipation and entanglement},\ }\href {https://doi.org/10.1103/PhysRevLett.88.097905} {\bibfield  {journal} {\bibinfo  {journal} {Phys. Rev. Lett.}\ }\textbf {\bibinfo {volume} {88}},\ \bibinfo {pages} {097905} (\bibinfo {year} {2002})}\BibitemShut {NoStop}%
\bibitem [{\citenamefont {Ziman}\ \emph {et~al.}(2001)\citenamefont {Ziman}, \citenamefont {\ifmmode \check{S}\else \v{S}\fi{}telmachovi\ifmmode~\check{c}\else \v{c}\fi{}}, \citenamefont {Buzek}, \citenamefont {Hillery}, \citenamefont {Scarani},\ and\ \citenamefont {Gisin}}]{ziman2001quantumhomogenization}%
  \BibitemOpen
  \bibfield  {author} {\bibinfo {author} {\bibfnamefont {M.}~\bibnamefont {Ziman}}, \bibinfo {author} {\bibfnamefont {P.}~\bibnamefont {\ifmmode \check{S}\else \v{S}\fi{}telmachovi\ifmmode~\check{c}\else \v{c}\fi{}}}, \bibinfo {author} {\bibfnamefont {V.}~\bibnamefont {Buzek}}, \bibinfo {author} {\bibfnamefont {M.}~\bibnamefont {Hillery}}, \bibinfo {author} {\bibfnamefont {V.}~\bibnamefont {Scarani}},\ and\ \bibinfo {author} {\bibfnamefont {N.}~\bibnamefont {Gisin}},\ }\href {https://arxiv.org/abs/quant-ph/0110164} {\bibinfo {title} {Quantum homogenization}} (\bibinfo {year} {2001}),\ \Eprint {https://arxiv.org/abs/quant-ph/0110164} {arXiv:quant-ph/0110164 [quant-ph]} \BibitemShut {NoStop}%
\bibitem [{\citenamefont {Rempe}\ \emph {et~al.}(1990)\citenamefont {Rempe}, \citenamefont {Schmidt-Kaler},\ and\ \citenamefont {Walther}}]{PhysRevLett.64.2783}%
  \BibitemOpen
  \bibfield  {author} {\bibinfo {author} {\bibfnamefont {G.}~\bibnamefont {Rempe}}, \bibinfo {author} {\bibfnamefont {F.}~\bibnamefont {Schmidt-Kaler}},\ and\ \bibinfo {author} {\bibfnamefont {H.}~\bibnamefont {Walther}},\ }\bibfield  {title} {\bibinfo {title} {Observation of sub-{P}oissonian photon statistics in a micromaser},\ }\href {https://doi.org/10.1103/PhysRevLett.64.2783} {\bibfield  {journal} {\bibinfo  {journal} {Phys. Rev. Lett.}\ }\textbf {\bibinfo {volume} {64}},\ \bibinfo {pages} {2783} (\bibinfo {year} {1990})}\BibitemShut {NoStop}%
\bibitem [{\citenamefont {Pielawa}\ \emph {et~al.}(2007)\citenamefont {Pielawa}, \citenamefont {Morigi}, \citenamefont {Vitali},\ and\ \citenamefont {Davidovich}}]{PhysRevLett.98.240401}%
  \BibitemOpen
  \bibfield  {author} {\bibinfo {author} {\bibfnamefont {S.}~\bibnamefont {Pielawa}}, \bibinfo {author} {\bibfnamefont {G.}~\bibnamefont {Morigi}}, \bibinfo {author} {\bibfnamefont {D.}~\bibnamefont {Vitali}},\ and\ \bibinfo {author} {\bibfnamefont {L.}~\bibnamefont {Davidovich}},\ }\bibfield  {title} {\bibinfo {title} {Generation of {E}instein-{P}odolsky-{R}osen-entangled radiation through an atomic reservoir},\ }\href {https://doi.org/10.1103/PhysRevLett.98.240401} {\bibfield  {journal} {\bibinfo  {journal} {Phys. Rev. Lett.}\ }\textbf {\bibinfo {volume} {98}},\ \bibinfo {pages} {240401} (\bibinfo {year} {2007})}\BibitemShut {NoStop}%
\bibitem [{\citenamefont {Miao}\ and\ \citenamefont {Sarlette}(2017)}]{Miao_2017}%
  \BibitemOpen
  \bibfield  {author} {\bibinfo {author} {\bibfnamefont {Z.}~\bibnamefont {Miao}}\ and\ \bibinfo {author} {\bibfnamefont {A.}~\bibnamefont {Sarlette}},\ }\bibfield  {title} {\bibinfo {title} {Discrete-time reservoir engineering with entangled bath and stabilising squeezed states},\ }\href {https://doi.org/10.1088/2058-9565/aa7ce8} {\bibfield  {journal} {\bibinfo  {journal} {Quantum Science and Technology}\ }\textbf {\bibinfo {volume} {2}},\ \bibinfo {pages} {034013} (\bibinfo {year} {2017})}\BibitemShut {NoStop}%
\bibitem [{\citenamefont {Goldberg}\ and\ \citenamefont {Holland}(1988)}]{Goldberg1988}%
  \BibitemOpen
  \bibfield  {author} {\bibinfo {author} {\bibfnamefont {D.~E.}\ \bibnamefont {Goldberg}}\ and\ \bibinfo {author} {\bibfnamefont {J.~H.}\ \bibnamefont {Holland}},\ }\bibfield  {title} {\bibinfo {title} {Genetic algorithms and machine learning},\ }\href {https://doi.org/10.1023/A:1022602019183} {\bibfield  {journal} {\bibinfo  {journal} {Machine Learning}\ }\textbf {\bibinfo {volume} {3}},\ \bibinfo {pages} {95} (\bibinfo {year} {1988})}\BibitemShut {NoStop}%
\bibitem [{\citenamefont {Holland}(1975)}]{holland1975adaptation}%
  \BibitemOpen
  \bibfield  {author} {\bibinfo {author} {\bibfnamefont {J.~H.}\ \bibnamefont {Holland}},\ }\href@noop {} {\emph {\bibinfo {title} {Adaptation in Natural and Artificial Systems}}}\ (\bibinfo  {publisher} {University of Michigan Press},\ \bibinfo {year} {1975})\BibitemShut {NoStop}%
\bibitem [{\citenamefont {Eiben}\ and\ \citenamefont {Schoenauer}(2005)}]{eiben2005evolutionarycomputing}%
  \BibitemOpen
  \bibfield  {author} {\bibinfo {author} {\bibfnamefont {A.~E.}\ \bibnamefont {Eiben}}\ and\ \bibinfo {author} {\bibfnamefont {M.}~\bibnamefont {Schoenauer}},\ }\href {https://arxiv.org/abs/cs/0511004} {\bibinfo {title} {Evolutionary computing}} (\bibinfo {year} {2005}),\ \Eprint {https://arxiv.org/abs/cs/0511004} {arXiv:cs/0511004 [cs.AI]} \BibitemShut {NoStop}%
\bibitem [{\citenamefont {Lukac}\ and\ \citenamefont {Perkowski}(2002)}]{lukac2002evolving}%
  \BibitemOpen
  \bibfield  {author} {\bibinfo {author} {\bibfnamefont {M.}~\bibnamefont {Lukac}}\ and\ \bibinfo {author} {\bibfnamefont {M.}~\bibnamefont {Perkowski}},\ }\bibfield  {title} {\bibinfo {title} {Evolving quantum circuits using genetic algorithm},\ }in\ \href@noop {} {\emph {\bibinfo {booktitle} {Proceedings 2002 NASA/DoD Conference on Evolvable Hardware}}}\ (\bibinfo {organization} {IEEE},\ \bibinfo {year} {2002})\ pp.\ \bibinfo {pages} {177--185}\BibitemShut {NoStop}%
\bibitem [{\citenamefont {Rubinstein}(2001)}]{rubinstein2001evolving}%
  \BibitemOpen
  \bibfield  {author} {\bibinfo {author} {\bibfnamefont {B.~I.}\ \bibnamefont {Rubinstein}},\ }\bibfield  {title} {\bibinfo {title} {Evolving quantum circuits using genetic programming},\ }in\ \href@noop {} {\emph {\bibinfo {booktitle} {Proceedings of the 2001 congress on evolutionary computation (IEEE Cat. No. 01TH8546)}}},\ Vol.~\bibinfo {volume} {1}\ (\bibinfo {organization} {IEEE},\ \bibinfo {year} {2001})\ pp.\ \bibinfo {pages} {144--151}\BibitemShut {NoStop}%
\bibitem [{\citenamefont {Williams}\ and\ \citenamefont {Gray}(1998)}]{williams1998automated}%
  \BibitemOpen
  \bibfield  {author} {\bibinfo {author} {\bibfnamefont {C.~P.}\ \bibnamefont {Williams}}\ and\ \bibinfo {author} {\bibfnamefont {A.~G.}\ \bibnamefont {Gray}},\ }\bibfield  {title} {\bibinfo {title} {Automated design of quantum circuits},\ }in\ \href@noop {} {\emph {\bibinfo {booktitle} {NASA International Conference on Quantum Computing and Quantum Communications}}}\ (\bibinfo {organization} {Springer},\ \bibinfo {year} {1998})\ pp.\ \bibinfo {pages} {113--125}\BibitemShut {NoStop}%
\bibitem [{\citenamefont {S{\"u}nkel}\ \emph {et~al.}(2023)\citenamefont {S{\"u}nkel}, \citenamefont {Martyniuk}, \citenamefont {Mattern}, \citenamefont {Jung},\ and\ \citenamefont {Paschke}}]{sunkel2023ga4qcogeneticalgorithmquantum}%
  \BibitemOpen
  \bibfield  {author} {\bibinfo {author} {\bibfnamefont {L.}~\bibnamefont {S{\"u}nkel}}, \bibinfo {author} {\bibfnamefont {D.}~\bibnamefont {Martyniuk}}, \bibinfo {author} {\bibfnamefont {D.}~\bibnamefont {Mattern}}, \bibinfo {author} {\bibfnamefont {J.}~\bibnamefont {Jung}},\ and\ \bibinfo {author} {\bibfnamefont {A.}~\bibnamefont {Paschke}},\ }\href {https://arxiv.org/abs/2302.01303} {\bibinfo {title} {Ga4qco: Genetic algorithm for quantum circuit optimization}} (\bibinfo {year} {2023}),\ \Eprint {https://arxiv.org/abs/2302.01303} {arXiv:2302.01303 [quant-ph]} \BibitemShut {NoStop}%
\bibitem [{\citenamefont {Brown}\ \emph {et~al.}(2023)\citenamefont {Brown}, \citenamefont {Paternostro},\ and\ \citenamefont {Ferraro}}]{Brown_2023}%
  \BibitemOpen
  \bibfield  {author} {\bibinfo {author} {\bibfnamefont {J.}~\bibnamefont {Brown}}, \bibinfo {author} {\bibfnamefont {M.}~\bibnamefont {Paternostro}},\ and\ \bibinfo {author} {\bibfnamefont {A.}~\bibnamefont {Ferraro}},\ }\bibfield  {title} {\bibinfo {title} {Optimal quantum control via genetic algorithms for quantum state engineering in driven-resonator mediated networks},\ }\href {https://doi.org/10.1088/2058-9565/acb2f2} {\bibfield  {journal} {\bibinfo  {journal} {Quantum Science and Technology}\ }\textbf {\bibinfo {volume} {8}},\ \bibinfo {pages} {025004} (\bibinfo {year} {2023})}\BibitemShut {NoStop}%
\bibitem [{\citenamefont {Palittapongarnpim}\ \emph {et~al.}(2017)\citenamefont {Palittapongarnpim}, \citenamefont {Wittek}, \citenamefont {Zahedinejad}, \citenamefont {Vedaie},\ and\ \citenamefont {Sanders}}]{PALITTAPONGARNPIM2017116}%
  \BibitemOpen
  \bibfield  {author} {\bibinfo {author} {\bibfnamefont {P.}~\bibnamefont {Palittapongarnpim}}, \bibinfo {author} {\bibfnamefont {P.}~\bibnamefont {Wittek}}, \bibinfo {author} {\bibfnamefont {E.}~\bibnamefont {Zahedinejad}}, \bibinfo {author} {\bibfnamefont {S.}~\bibnamefont {Vedaie}},\ and\ \bibinfo {author} {\bibfnamefont {B.~C.}\ \bibnamefont {Sanders}},\ }\bibfield  {title} {\bibinfo {title} {Learning in quantum control: High-dimensional global optimization for noisy quantum dynamics},\ }\href {https://doi.org/https://doi.org/10.1016/j.neucom.2016.12.087} {\bibfield  {journal} {\bibinfo  {journal} {Neurocomputing}\ }\textbf {\bibinfo {volume} {268}},\ \bibinfo {pages} {116} (\bibinfo {year} {2017})}\BibitemShut {NoStop}%
\bibitem [{\citenamefont {Zahedinejad}\ \emph {et~al.}(2014)\citenamefont {Zahedinejad}, \citenamefont {Schirmer},\ and\ \citenamefont {Sanders}}]{PhysRevA.90.032310}%
  \BibitemOpen
  \bibfield  {author} {\bibinfo {author} {\bibfnamefont {E.}~\bibnamefont {Zahedinejad}}, \bibinfo {author} {\bibfnamefont {S.}~\bibnamefont {Schirmer}},\ and\ \bibinfo {author} {\bibfnamefont {B.~C.}\ \bibnamefont {Sanders}},\ }\bibfield  {title} {\bibinfo {title} {Evolutionary algorithms for hard quantum control},\ }\href {https://doi.org/10.1103/PhysRevA.90.032310} {\bibfield  {journal} {\bibinfo  {journal} {Phys. Rev. A}\ }\textbf {\bibinfo {volume} {90}},\ \bibinfo {pages} {032310} (\bibinfo {year} {2014})}\BibitemShut {NoStop}%
\bibitem [{\citenamefont {Chisholm}\ \emph {et~al.}(2021)\citenamefont {Chisholm}, \citenamefont {Garc{\'i}a-P{\'e}rez}, \citenamefont {Rossi}, \citenamefont {Palma},\ and\ \citenamefont {Maniscalco}}]{Chisholm_2021}%
  \BibitemOpen
  \bibfield  {author} {\bibinfo {author} {\bibfnamefont {D.~A.}\ \bibnamefont {Chisholm}}, \bibinfo {author} {\bibfnamefont {G.}~\bibnamefont {Garc{\'i}a-P{\'e}rez}}, \bibinfo {author} {\bibfnamefont {M.~A.~C.}\ \bibnamefont {Rossi}}, \bibinfo {author} {\bibfnamefont {G.~M.}\ \bibnamefont {Palma}},\ and\ \bibinfo {author} {\bibfnamefont {S.}~\bibnamefont {Maniscalco}},\ }\bibfield  {title} {\bibinfo {title} {Stochastic collision model approach to transport phenomena in quantum networks},\ }\href {https://doi.org/10.1088/1367-2630/abd57d} {\bibfield  {journal} {\bibinfo  {journal} {New Journal of Physics}\ }\textbf {\bibinfo {volume} {23}},\ \bibinfo {pages} {033031} (\bibinfo {year} {2021})}\BibitemShut {NoStop}%
\bibitem [{\citenamefont {Raimond}\ and\ \citenamefont {Haroche}(2006)}]{raimond2006exploring}%
  \BibitemOpen
  \bibfield  {author} {\bibinfo {author} {\bibfnamefont {J.-M.}\ \bibnamefont {Raimond}}\ and\ \bibinfo {author} {\bibfnamefont {S.}~\bibnamefont {Haroche}},\ }\bibfield  {title} {\bibinfo {title} {Exploring the quantum},\ }\href@noop {} {\bibfield  {journal} {\bibinfo  {journal} {Oxford University Press}\ }\textbf {\bibinfo {volume} {82}},\ \bibinfo {pages} {17} (\bibinfo {year} {2006})}\BibitemShut {NoStop}%
\bibitem [{\citenamefont {Konak}\ \emph {et~al.}(2006)\citenamefont {Konak}, \citenamefont {Coit},\ and\ \citenamefont {Smith}}]{KONAK2006992}%
  \BibitemOpen
  \bibfield  {author} {\bibinfo {author} {\bibfnamefont {A.}~\bibnamefont {Konak}}, \bibinfo {author} {\bibfnamefont {D.~W.}\ \bibnamefont {Coit}},\ and\ \bibinfo {author} {\bibfnamefont {A.~E.}\ \bibnamefont {Smith}},\ }\bibfield  {title} {\bibinfo {title} {Multi-objective optimization using genetic algorithms: A tutorial},\ }\href {https://doi.org/https://doi.org/10.1016/j.ress.2005.11.018} {\bibfield  {journal} {\bibinfo  {journal} {Reliability Engineering \& System Safety}\ }\textbf {\bibinfo {volume} {91}},\ \bibinfo {pages} {992} (\bibinfo {year} {2006})}\BibitemShut {NoStop}%
\bibitem [{\citenamefont {Nielsen}\ and\ \citenamefont {Chuang}(2010)}]{Nielsen_Chuang_2010}%
  \BibitemOpen
  \bibfield  {author} {\bibinfo {author} {\bibfnamefont {M.~A.}\ \bibnamefont {Nielsen}}\ and\ \bibinfo {author} {\bibfnamefont {I.~L.}\ \bibnamefont {Chuang}},\ }\href@noop {} {\emph {\bibinfo {title} {Quantum Computation and Quantum Information: 10th Anniversary Edition}}}\ (\bibinfo  {publisher} {Cambridge University Press},\ \bibinfo {year} {2010})\BibitemShut {NoStop}%
\bibitem [{\citenamefont {Tilma}\ and\ \citenamefont {Sudarshan}(2002)}]{ToddTilma_2002}%
  \BibitemOpen
  \bibfield  {author} {\bibinfo {author} {\bibfnamefont {T.}~\bibnamefont {Tilma}}\ and\ \bibinfo {author} {\bibfnamefont {E.~C.~G.}\ \bibnamefont {Sudarshan}},\ }\bibfield  {title} {\bibinfo {title} {Generalized {E}uler angle parametrization for {SU(N)}},\ }\href {https://doi.org/10.1088/0305-4470/35/48/316} {\bibfield  {journal} {\bibinfo  {journal} {Journal of Physics A: Mathematical and General}\ }\textbf {\bibinfo {volume} {35}},\ \bibinfo {pages} {10467} (\bibinfo {year} {2002})}\BibitemShut {NoStop}%
\bibitem [{\citenamefont {Tilma}\ \emph {et~al.}(2002)\citenamefont {Tilma}, \citenamefont {Byrd},\ and\ \citenamefont {Sudarshan}}]{ToddTilma_2002_2}%
  \BibitemOpen
  \bibfield  {author} {\bibinfo {author} {\bibfnamefont {T.}~\bibnamefont {Tilma}}, \bibinfo {author} {\bibfnamefont {M.}~\bibnamefont {Byrd}},\ and\ \bibinfo {author} {\bibfnamefont {E.~C.~G.}\ \bibnamefont {Sudarshan}},\ }\bibfield  {title} {\bibinfo {title} {A parametrization of bipartite systems based on {SU}(4) {E}uler angles},\ }\href {https://doi.org/10.1088/0305-4470/35/48/315} {\bibfield  {journal} {\bibinfo  {journal} {Journal of Physics A: Mathematical and General}\ }\textbf {\bibinfo {volume} {35}},\ \bibinfo {pages} {10445} (\bibinfo {year} {2002})}\BibitemShut {NoStop}%
\bibitem [{\citenamefont {Byrd}(1997)}]{byrd1997geometrysu3}%
  \BibitemOpen
  \bibfield  {author} {\bibinfo {author} {\bibfnamefont {M.}~\bibnamefont {Byrd}},\ }\href {https://arxiv.org/abs/physics/9708015} {\bibinfo {title} {The geometry of su(3)}} (\bibinfo {year} {1997}),\ \Eprint {https://arxiv.org/abs/physics/9708015} {arXiv:physics/9708015 [math-ph]} \BibitemShut {NoStop}%
\bibitem [{\citenamefont {Haupt}\ and\ \citenamefont {Haupt}(2004)}]{haupt2004practical}%
  \BibitemOpen
  \bibfield  {author} {\bibinfo {author} {\bibfnamefont {R.~L.}\ \bibnamefont {Haupt}}\ and\ \bibinfo {author} {\bibfnamefont {S.~E.}\ \bibnamefont {Haupt}},\ }\href@noop {} {\emph {\bibinfo {title} {Practical genetic algorithms}}}\ (\bibinfo  {publisher} {John Wiley \& Sons},\ \bibinfo {year} {2004})\BibitemShut {NoStop}%
\bibitem [{\citenamefont {Rodrigues}\ \emph {et~al.}(2019)\citenamefont {Rodrigues}, \citenamefont {De~Chiara}, \citenamefont {Paternostro},\ and\ \citenamefont {Landi}}]{PhysRevLett.123.140601}%
  \BibitemOpen
  \bibfield  {author} {\bibinfo {author} {\bibfnamefont {F.~L.~S.}\ \bibnamefont {Rodrigues}}, \bibinfo {author} {\bibfnamefont {G.}~\bibnamefont {De~Chiara}}, \bibinfo {author} {\bibfnamefont {M.}~\bibnamefont {Paternostro}},\ and\ \bibinfo {author} {\bibfnamefont {G.~T.}\ \bibnamefont {Landi}},\ }\bibfield  {title} {\bibinfo {title} {Thermodynamics of weakly coherent collisional models},\ }\href {https://doi.org/10.1103/PhysRevLett.123.140601} {\bibfield  {journal} {\bibinfo  {journal} {Phys. Rev. Lett.}\ }\textbf {\bibinfo {volume} {123}},\ \bibinfo {pages} {140601} (\bibinfo {year} {2019})}\BibitemShut {NoStop}%
\bibitem [{\citenamefont {Loudon}(2000)}]{loudon2000quantum}%
  \BibitemOpen
  \bibfield  {author} {\bibinfo {author} {\bibfnamefont {R.}~\bibnamefont {Loudon}},\ }\href@noop {} {\emph {\bibinfo {title} {The quantum theory of light}}}\ (\bibinfo  {publisher} {OUP Oxford},\ \bibinfo {year} {2000})\BibitemShut {NoStop}%
\bibitem [{\citenamefont {Walls}(1983)}]{Walls1983}%
  \BibitemOpen
  \bibfield  {author} {\bibinfo {author} {\bibfnamefont {D.~F.}\ \bibnamefont {Walls}},\ }\bibfield  {title} {\bibinfo {title} {Squeezed states of light},\ }\href {https://doi.org/10.1038/306141a0} {\bibfield  {journal} {\bibinfo  {journal} {Nature}\ }\textbf {\bibinfo {volume} {306}},\ \bibinfo {pages} {141} (\bibinfo {year} {1983})}\BibitemShut {NoStop}%
\bibitem [{\citenamefont {Eisert}\ \emph {et~al.}(2002)\citenamefont {Eisert}, \citenamefont {Scheel},\ and\ \citenamefont {Plenio}}]{PhysRevLett.89.137903}%
  \BibitemOpen
  \bibfield  {author} {\bibinfo {author} {\bibfnamefont {J.}~\bibnamefont {Eisert}}, \bibinfo {author} {\bibfnamefont {S.}~\bibnamefont {Scheel}},\ and\ \bibinfo {author} {\bibfnamefont {M.~B.}\ \bibnamefont {Plenio}},\ }\bibfield  {title} {\bibinfo {title} {Distilling {G}aussian states with {G}aussian operations is impossible},\ }\href {https://doi.org/10.1103/PhysRevLett.89.137903} {\bibfield  {journal} {\bibinfo  {journal} {Phys. Rev. Lett.}\ }\textbf {\bibinfo {volume} {89}},\ \bibinfo {pages} {137903} (\bibinfo {year} {2002})}\BibitemShut {NoStop}%
\bibitem [{\citenamefont {Niset}\ \emph {et~al.}(2009)\citenamefont {Niset}, \citenamefont {Fiur\'a\ifmmode~\check{s}\else \v{s}\fi{}ek},\ and\ \citenamefont {Cerf}}]{PhysRevLett.102.120501}%
  \BibitemOpen
  \bibfield  {author} {\bibinfo {author} {\bibfnamefont {J.}~\bibnamefont {Niset}}, \bibinfo {author} {\bibfnamefont {J.}~\bibnamefont {Fiur\'a\ifmmode~\check{s}\else \v{s}\fi{}ek}},\ and\ \bibinfo {author} {\bibfnamefont {N.~J.}\ \bibnamefont {Cerf}},\ }\bibfield  {title} {\bibinfo {title} {No-go theorem for {G}aussian quantum error correction},\ }\href {https://doi.org/10.1103/PhysRevLett.102.120501} {\bibfield  {journal} {\bibinfo  {journal} {Phys. Rev. Lett.}\ }\textbf {\bibinfo {volume} {102}},\ \bibinfo {pages} {120501} (\bibinfo {year} {2009})}\BibitemShut {NoStop}%
\bibitem [{\citenamefont {Ferraro}\ and\ \citenamefont {Paris}(2005)}]{Ferraro_2005}%
  \BibitemOpen
  \bibfield  {author} {\bibinfo {author} {\bibfnamefont {A.}~\bibnamefont {Ferraro}}\ and\ \bibinfo {author} {\bibfnamefont {M.~G.~A.}\ \bibnamefont {Paris}},\ }\bibfield  {title} {\bibinfo {title} {Nonlocality of two- and three-mode continuous variable systems},\ }\href {https://doi.org/10.1088/1464-4266/7/6/003} {\bibfield  {journal} {\bibinfo  {journal} {Journal of Optics B: Quantum and Semiclassical Optics}\ }\textbf {\bibinfo {volume} {7}},\ \bibinfo {pages} {174–182} (\bibinfo {year} {2005})}\BibitemShut {NoStop}%
\bibitem [{\citenamefont {Genoni}\ and\ \citenamefont {Paris}(2010)}]{PhysRevA.82.052341}%
  \BibitemOpen
  \bibfield  {author} {\bibinfo {author} {\bibfnamefont {M.~G.}\ \bibnamefont {Genoni}}\ and\ \bibinfo {author} {\bibfnamefont {M.~G.~A.}\ \bibnamefont {Paris}},\ }\bibfield  {title} {\bibinfo {title} {Quantifying non-{G}aussianity for quantum information},\ }\href {https://doi.org/10.1103/PhysRevA.82.052341} {\bibfield  {journal} {\bibinfo  {journal} {Phys. Rev. A}\ }\textbf {\bibinfo {volume} {82}},\ \bibinfo {pages} {052341} (\bibinfo {year} {2010})}\BibitemShut {NoStop}%
\bibitem [{\citenamefont {Harris}\ \emph {et~al.}(2020)\citenamefont {Harris}, \citenamefont {Millman}, \citenamefont {van~der Walt}, \citenamefont {Gommers}, \citenamefont {Virtanen}, \citenamefont {Cournapeau}, \citenamefont {Wieser}, \citenamefont {Taylor}, \citenamefont {Berg}, \citenamefont {Smith}, \citenamefont {Kern}, \citenamefont {Picus}, \citenamefont {Hoyer}, \citenamefont {van Kerkwijk}, \citenamefont {Brett}, \citenamefont {Haldane}, \citenamefont {del R{\'{i}}o}, \citenamefont {Wiebe}, \citenamefont {Peterson}, \citenamefont {G{\'{e}}rard-Marchant}, \citenamefont {Sheppard}, \citenamefont {Reddy}, \citenamefont {Weckesser}, \citenamefont {Abbasi}, \citenamefont {Gohlke},\ and\ \citenamefont {Oliphant}}]{numpybib}%
  \BibitemOpen
  \bibfield  {author} {\bibinfo {author} {\bibfnamefont {C.~R.}\ \bibnamefont {Harris}}, \bibinfo {author} {\bibfnamefont {K.~J.}\ \bibnamefont {Millman}}, \bibinfo {author} {\bibfnamefont {S.~J.}\ \bibnamefont {van~der Walt}}, \bibinfo {author} {\bibfnamefont {R.}~\bibnamefont {Gommers}}, \bibinfo {author} {\bibfnamefont {P.}~\bibnamefont {Virtanen}}, \bibinfo {author} {\bibfnamefont {D.}~\bibnamefont {Cournapeau}}, \bibinfo {author} {\bibfnamefont {E.}~\bibnamefont {Wieser}}, \bibinfo {author} {\bibfnamefont {J.}~\bibnamefont {Taylor}}, \bibinfo {author} {\bibfnamefont {S.}~\bibnamefont {Berg}}, \bibinfo {author} {\bibfnamefont {N.~J.}\ \bibnamefont {Smith}}, \bibinfo {author} {\bibfnamefont {R.}~\bibnamefont {Kern}}, \bibinfo {author} {\bibfnamefont {M.}~\bibnamefont {Picus}}, \bibinfo {author} {\bibfnamefont {S.}~\bibnamefont {Hoyer}}, \bibinfo {author} {\bibfnamefont {M.~H.}\ \bibnamefont {van Kerkwijk}}, \bibinfo {author} {\bibfnamefont {M.}~\bibnamefont {Brett}}, \bibinfo {author} {\bibfnamefont
  {A.}~\bibnamefont {Haldane}}, \bibinfo {author} {\bibfnamefont {J.~F.}\ \bibnamefont {del R{\'{i}}o}}, \bibinfo {author} {\bibfnamefont {M.}~\bibnamefont {Wiebe}}, \bibinfo {author} {\bibfnamefont {P.}~\bibnamefont {Peterson}}, \bibinfo {author} {\bibfnamefont {P.}~\bibnamefont {G{\'{e}}rard-Marchant}}, \bibinfo {author} {\bibfnamefont {K.}~\bibnamefont {Sheppard}}, \bibinfo {author} {\bibfnamefont {T.}~\bibnamefont {Reddy}}, \bibinfo {author} {\bibfnamefont {W.}~\bibnamefont {Weckesser}}, \bibinfo {author} {\bibfnamefont {H.}~\bibnamefont {Abbasi}}, \bibinfo {author} {\bibfnamefont {C.}~\bibnamefont {Gohlke}},\ and\ \bibinfo {author} {\bibfnamefont {T.~E.}\ \bibnamefont {Oliphant}},\ }\bibfield  {title} {\bibinfo {title} {Array programming with {NumPy}},\ }\href {https://doi.org/10.1038/s41586-020-2649-2} {\bibfield  {journal} {\bibinfo  {journal} {Nature}\ }\textbf {\bibinfo {volume} {585}},\ \bibinfo {pages} {357} (\bibinfo {year} {2020})}\BibitemShut {NoStop}%
\bibitem [{\citenamefont {Virtanen}\ \emph {et~al.}(2020)\citenamefont {Virtanen}, \citenamefont {Gommers}, \citenamefont {Oliphant}, \citenamefont {Haberland}, \citenamefont {Reddy}, \citenamefont {Cournapeau}, \citenamefont {Burovski}, \citenamefont {Peterson}, \citenamefont {Weckesser}, \citenamefont {Bright}, \citenamefont {{van der Walt}}, \citenamefont {Brett}, \citenamefont {Wilson}, \citenamefont {Millman}, \citenamefont {Mayorov}, \citenamefont {Nelson}, \citenamefont {Jones}, \citenamefont {Kern}, \citenamefont {Larson}, \citenamefont {Carey}, \citenamefont {Polat}, \citenamefont {Feng}, \citenamefont {Moore}, \citenamefont {{VanderPlas}}, \citenamefont {Laxalde}, \citenamefont {Perktold}, \citenamefont {Cimrman}, \citenamefont {Henriksen}, \citenamefont {Quintero}, \citenamefont {Harris}, \citenamefont {Archibald}, \citenamefont {Ribeiro}, \citenamefont {Pedregosa}, \citenamefont {{van Mulbregt}},\ and\ \citenamefont {{SciPy 1.0 Contributors}}}]{scipybib}%
  \BibitemOpen
  \bibfield  {author} {\bibinfo {author} {\bibfnamefont {P.}~\bibnamefont {Virtanen}}, \bibinfo {author} {\bibfnamefont {R.}~\bibnamefont {Gommers}}, \bibinfo {author} {\bibfnamefont {T.~E.}\ \bibnamefont {Oliphant}}, \bibinfo {author} {\bibfnamefont {M.}~\bibnamefont {Haberland}}, \bibinfo {author} {\bibfnamefont {T.}~\bibnamefont {Reddy}}, \bibinfo {author} {\bibfnamefont {D.}~\bibnamefont {Cournapeau}}, \bibinfo {author} {\bibfnamefont {E.}~\bibnamefont {Burovski}}, \bibinfo {author} {\bibfnamefont {P.}~\bibnamefont {Peterson}}, \bibinfo {author} {\bibfnamefont {W.}~\bibnamefont {Weckesser}}, \bibinfo {author} {\bibfnamefont {J.}~\bibnamefont {Bright}}, \bibinfo {author} {\bibfnamefont {S.~J.}\ \bibnamefont {{van der Walt}}}, \bibinfo {author} {\bibfnamefont {M.}~\bibnamefont {Brett}}, \bibinfo {author} {\bibfnamefont {J.}~\bibnamefont {Wilson}}, \bibinfo {author} {\bibfnamefont {K.~J.}\ \bibnamefont {Millman}}, \bibinfo {author} {\bibfnamefont {N.}~\bibnamefont {Mayorov}}, \bibinfo {author} {\bibfnamefont
  {A.~R.~J.}\ \bibnamefont {Nelson}}, \bibinfo {author} {\bibfnamefont {E.}~\bibnamefont {Jones}}, \bibinfo {author} {\bibfnamefont {R.}~\bibnamefont {Kern}}, \bibinfo {author} {\bibfnamefont {E.}~\bibnamefont {Larson}}, \bibinfo {author} {\bibfnamefont {C.~J.}\ \bibnamefont {Carey}}, \bibinfo {author} {\bibfnamefont {{\.I}.}~\bibnamefont {Polat}}, \bibinfo {author} {\bibfnamefont {Y.}~\bibnamefont {Feng}}, \bibinfo {author} {\bibfnamefont {E.~W.}\ \bibnamefont {Moore}}, \bibinfo {author} {\bibfnamefont {J.}~\bibnamefont {{VanderPlas}}}, \bibinfo {author} {\bibfnamefont {D.}~\bibnamefont {Laxalde}}, \bibinfo {author} {\bibfnamefont {J.}~\bibnamefont {Perktold}}, \bibinfo {author} {\bibfnamefont {R.}~\bibnamefont {Cimrman}}, \bibinfo {author} {\bibfnamefont {I.}~\bibnamefont {Henriksen}}, \bibinfo {author} {\bibfnamefont {E.~A.}\ \bibnamefont {Quintero}}, \bibinfo {author} {\bibfnamefont {C.~R.}\ \bibnamefont {Harris}}, \bibinfo {author} {\bibfnamefont {A.~M.}\ \bibnamefont {Archibald}}, \bibinfo {author}
  {\bibfnamefont {A.~H.}\ \bibnamefont {Ribeiro}}, \bibinfo {author} {\bibfnamefont {F.}~\bibnamefont {Pedregosa}}, \bibinfo {author} {\bibfnamefont {P.}~\bibnamefont {{van Mulbregt}}},\ and\ \bibinfo {author} {\bibnamefont {{SciPy 1.0 Contributors}}},\ }\bibfield  {title} {\bibinfo {title} {{{SciPy} 1.0: Fundamental Algorithms for Scientific Computing in Python}},\ }\href {https://doi.org/10.1038/s41592-019-0686-2} {\bibfield  {journal} {\bibinfo  {journal} {Nature Methods}\ }\textbf {\bibinfo {volume} {17}},\ \bibinfo {pages} {261} (\bibinfo {year} {2020})}\BibitemShut {NoStop}%
\end{thebibliography}%

\appendix
\section{Ancillae parametrization}\label{appendix1}
Here, we describe the parametrization of a set of system-ancilla physical parameters and ancilla states as a candidate solution of the genetic algorithms in Section \ref{section:geneticalgorithms} in more detail. As already explained, each candidate solution is expressed as a string $x$ whose components are real number $x_i\in[0,1]$. The first $\lambda_0$ components represent the, constant, rescaled, system-ancilla parameters, the true value of which are obtained from a linear transformation $[0,1]\rightarrow[X^m_i, X^M_i]$
\begin{equation}\label{equation:lineartransform}
    X_i = X^m_i + (X^M_i -X^m_i) x_i.
\end{equation}
We are now left with the parametrization of the ancilla states. Each state is described by $\lambda_d$ components, depending on the ancilla type and parametrization. 

For diagonal qubits, $\lambda_d =1$. A single parameter, representing a generalized temperature, is sufficient to describe the ancilla state. A maximum and a minimum ancilla inverse temperatures $\beta_A^{m}, \beta_A^{M}$ are fixed as hyperparameters. After applying Eq.~\eqref{equation:lineartransform} to $x_i$ with $X^m_i, X^M_i = \beta_A^{m}, \beta_A^{M}$, we obtain an inverse temperature $\beta_A^i$, from which we can construct
\begin{equation}
    \rho_A^{i} = \frac{e^{-\beta_A^i H_A}}{\tr(e^{-\beta_A^i H_A})}.
\end{equation}

For general qubits, we have $\lambda_d =3$ and each ancilla state is described by a Bloch vector. The components of such vector are written as $B_x^i=r^i \sin\theta^i\cos\phi^i$, $B_y^i=r^i \sin\theta^i\cos\phi^i$, $B_z^i=r^i\cos\theta^i$ where $r^i\in[0,1], \theta^i\in[0,\pi], \phi^i\in[0,2\pi]$ are obtained from the string entries using Eq.~\eqref{equation:lineartransform}. The ancilla density matrix is finally \cite{Nielsen_Chuang_2010}
\begin{equation}
    \rho_A^{i} = \frac{1}{2}(I + \sigma_X B_x^i + \sigma_Y B_y^i + \sigma_Z B_z^i),
\end{equation}
where $I$ is the identity and $\sigma_X, \sigma_Y, \sigma_Z$ the Pauli matrices.

A generic three level system state can be described by $\lambda_d =8$ real numbers. In order to avoid overparametrization while, at the same time, ensure a physically meaningful density matrix (i.e. positive semi-definite matrix with trace one) we deploy a parametrization based on Euler Angles \cite{ToddTilma_2002, ToddTilma_2002_2}. A generic density matrix can be obtained from an appropriate diagonal density matrix $\rho_D$ and a unitary transformation $U$. A generic unitary transformation in SU(3) can be written, in terms of Euler angles, as \cite{byrd1997geometrysu3}
\begin{equation}\label{equation:eulerU}
    U = e^{i\lambda_3\alpha}e^{i\lambda_2\beta}e^{i\lambda_3\gamma}e^{i\lambda_5\theta}e^{i\lambda_3a}e^{i\lambda_2b}e^{i\lambda_3c}e^{i\lambda_8\phi},
\end{equation}
where $\alpha,\gamma,a,c\in[0,\pi[$, $\beta, b, \theta\in[0,\pi/2]$, $\phi\in[0,2\pi[$ and $\lambda_i$ are the Gell-Mann matrices
\begin{equation*}
    \lambda_1=
    \begin{pmatrix}
        0 & 1 & 0\\
        1 & 0 & 0\\
        0 & 0 & 0
    \end{pmatrix},
    \lambda_2=
    \begin{pmatrix}
        0 & -i & 0\\
        i & 0 & 0\\
        0 & 0 & 0
    \end{pmatrix},
    \lambda_3=
    \begin{pmatrix}
        1 & 0 & 0\\
        0 & -1 & 0\\
        0 & 0 & 0
    \end{pmatrix},
\end{equation*}
\begin{equation*}
    \lambda_4=
    \begin{pmatrix}
        0 & 0 & 1\\
        0 & 0 & 0\\
        1 & 0 & 0
    \end{pmatrix},
    \lambda_5=
    \begin{pmatrix}
        0 & 0 & -i\\
        0 & 0 & 0\\
        i & 0 & 0
    \end{pmatrix},
    \lambda_6=
    \begin{pmatrix}
        0 & 0 & 0\\
        0 & 0 & 1\\
        0 & 1 & 0
    \end{pmatrix},
\end{equation*}
\begin{equation*}
    \lambda_7=
    \begin{pmatrix}
        0 & 0 & 0\\
        0 & 0 & -i\\
        0 & i & 0
    \end{pmatrix},
    \lambda_8=\frac{1}{\sqrt{3}}
    \begin{pmatrix}
        1 & 0 & 0\\
        0 & 0 & 0\\
        0 & 0 & -2
    \end{pmatrix}.
\end{equation*}
A generic diagonal $3\times3$ trace one matrix can be written as
\begin{equation*}
    \rho_D=
    \begin{pmatrix}
        z^2(1-y^2) & 0 & 0\\
        0 & z^2y^2 & 0\\
        0 & 0 & 1-y^2
    \end{pmatrix},
\end{equation*}
where $z^2=\sin^2\eta$ and $y^2=\sin^2\delta$. In terms of $\lambda_i$
\begin{equation}
    \rho_D =w_1I + w_2\lambda_3 +w_3\lambda_8,
\end{equation}
with $w_1=\frac{1}{3}(1+z^2-y^2)$, $w_2 =\frac{z^2}{2}-z^2y^2$ and $w_3=\frac{\sqrt{3}}{6}(z^2+2y^2-2)$. We want to write the ancilla state as $U\rho_DU^\dagger$, using Eq.~\eqref{equation:eulerU}. Noticing that $[I,\lambda_3]=[I,\lambda_8]=[\lambda_3, \lambda_8]=0$, we can simplify to
\begin{equation}\label{equation:Eulerho}
    \rho_A = V\rho_DV^\dagger,
\end{equation}
with
$V = e^{i\lambda_3\alpha}e^{i\lambda_2\beta}e^{i\lambda_3\gamma}e^{i\lambda_5\theta}e^{i\lambda_3a}e^{i\lambda_2b}$.
The strategy to parametrize $\rho_A^i$ is then to assign the candidate solution entries to the corresponding angles using Eq.~\eqref{equation:lineartransform} so that the density matrix can be constructed from Eq.~\eqref{equation:Eulerho}.

\section{Numerical and hyperparameter details}\label{appendix2}
All the numerical simulations have been performed using Python 3, in particular the modules NumPy \cite{numpybib} and SciPy \cite{scipybib}. 
The cavity field was approximated to the first $20$ energy levels in all cases except in Section \ref{section:results:squeezedstates} for $\zeta=1$ and in Section \ref{section:results:nongaussianstates}, where we considered the first $30$ levels. The range of allowed generalized inverse temperatures for diagonal ancilla states was $\beta_A\in[-5,5]$. The range of allowed coupling strengths was, in all cases, $g_l, g_{1l}, g_{2l}, g_{nl} \in [-1,1]$. The range of qubit frequencies was $\omega_A \in [0,5]$. For three level system ancillae, $\omega_1\in [0,5]$ and $\omega_2 =\omega_1 + \Delta \omega_{12}$, with $\Delta \omega_{12} \in [0,5]$. 
The mutation factor was fixed to $\mu=1$ and the number of extraction for tournament selection to $K=4$. For all applications of the algorithm described in Section \ref{section:geneticalgorithms:fixedlength} we set $N=200$, $M=100$ and $\nu_{steps} = 1000$ iterations. For applications of \ref{section:geneticalgorithms:variablelength} we set $N=200$, $M=50$ and $\bar{p} = 0.05$ while the maximum number of collision was $n_{max}=100$. The number of iterations was $\nu_{steps}\leq3000$ (where $\nu_{steps}=3000$ was used for squeezed state preparation with three level system ancillae).

\section{Flowchat of the algorithm}
We present the generic structure of a genetic algorithm in the form of the following flowchart  
\begin{figure}
\centering\includegraphics[width=0.6\columnwidth]{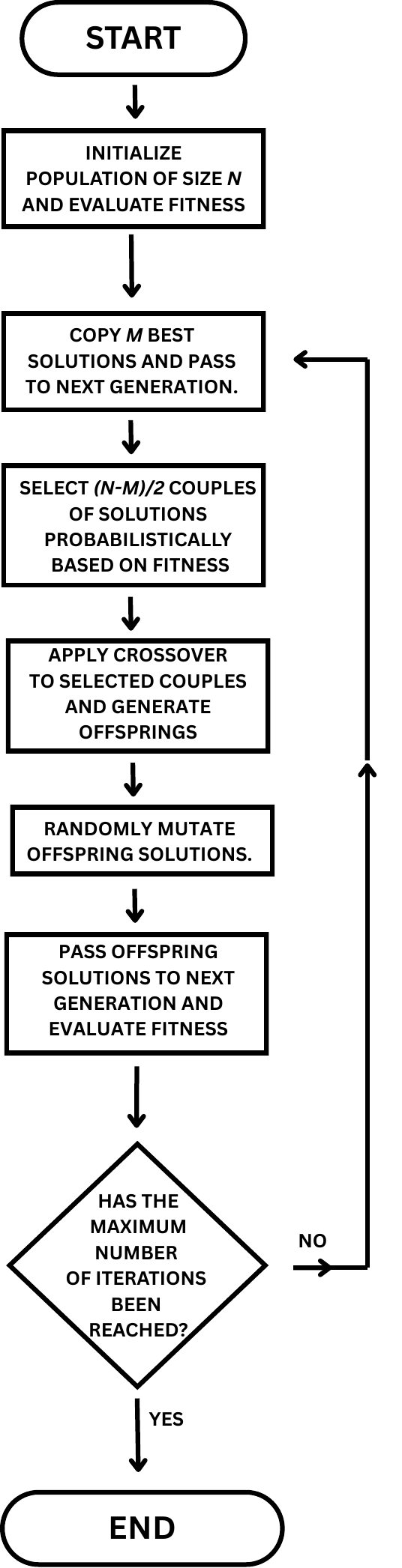}
\caption{Flowchart of the generic Genetic Algorithm described in Sec.~\ref{section:geneticalgorithms}.}\label{Figure:flowchart}
\end{figure}
\end{document}